\documentclass[pra,aps,twocolumn,amssymb,nofootinbib,showpacs,amsmath]{revtex4-1}
\usepackage{hyperref}
\usepackage{epsfig}
\begin{document}
\title{On the coherent effect of vacuum fluctuations on  driven atoms}
\author{M. Donaire  and A. Lambrecht}
\affiliation{Laboratoire Kastler Brossel, UPMC-Sorbonnes Universit\'es, ENS-PSL Research Universities, CNRS, Coll\`{e}ge de France, 4, place Jussieu, F-75252 Paris, France}
\date{10 April 2015}
\begin{abstract}
We study the coherent effect of  the Casimir-Polder interaction on the oscillations of two-photon driven atoms. We find that, for  oscillations between two degenerate states in lambda-configuration, shifts on the Rabi frequency may be induced by non-additive level shifts. For  oscillations between two Rydberg states in ladder-configuration, shifts on the Rabi frequency may be induced by the effective renormalization of the laser interaction. 

%
\end{abstract}
\maketitle

\section{Introduction}\label{intro}


The interaction of a neutral atom with a material surface is a  problem profusely addressed in the literature \cite{Casimir-Polder1948,Wylie,Buhmann,Gorza,Henkel,Safari}. In most of the approaches the atom is taken in a stationary state w.r.t. the time of observation. 
At zero temperature and in the electric dipole approximation,  the atom undergoes a series of virtual E1 transitions to upper levels. It is the coupling of the charges of the atom and the currents on the surface to the quantum EM field that induces the correlation between their transtient dipole moments, giving rise to a non-vanishing interaction. The lifetime of the virtual atomic transitions is very short in comparison to ordinary observation times and thus, the use of stationary quantum perturbation theory is well justified for the calculation of this interaction \cite{Craigbook}.
For distances greater than the relevant atomic transition wavelengths this interaction is referred to as \emph{retarded Casimir-Polder (CP) interaction}, while for distances much shorter than those wavelengths it is referred to as \emph{van der Waals} or \emph{non-retarded CP interaction}. 

When atoms close to dielectric surfaces are driven under the action of external sources, virtual and actual transitions mix up with each other in the time evolution of the atomic wave function. From a practical perspective, the effect of the CP interaction on the dynamics of driven atoms is of great importance in hybrid quantum systems --eg. Ref.\cite{chip}. At first sight, under quasi-stationary conditions, that effect reduces to an additive level shift on the atomic eigenstates, which is just a generalized Lamb-shift \cite{Wylie,Safari,PRADonaire2}. In the closely related case of the interaction between two driven Rydberg atoms, this phenomenon originates the van der Waals blockade of the Rabi oscillations. This is the idea behind neutral atoms quantum gates \cite{Jakschetal2000}, where the excitation of the target atom is blocked as the energy shift of its Rydberg level exceeds the value of the bare Rabi frequency. However, recent findings suggests that other dynamical effects might be relevant under certain conditions --cf. Refs.\cite{EPL,Ribeiro}.

On the other hand, state of the art techniques on atomic interferometry  have proved useful in measuring accurately the dynamical phase shifts accumulated by the wave function of coherently driven atoms \cite{PRLPelisson}. Those phase shifts contain information about the interaction of the atoms with the environment.
This is at the root  of the proposals of Refs.\cite{PRALambrechtetal,Sorrentino} to measure  the interaction of an alkali atom with a macroscopic surface at
submicron distances. At this distance, the dominant interactions are expected to be the CP interaction and the gravitational interaction. In the
experimental setup of Ref.\cite{PRALambrechtetal} the atoms are trapped in a 1D vertical lattice, which allows for an accurate control over the relative position of the atoms w.r.t. to the surface. The uniform Earth gravitational field creates a ladder of localized states referred to as Wannier-Stack states. At first glance, the net effect of the Casimir interaction reduces to additive shifts on the  atomic energy levels, both external (i.e., those involving the center of mass d.o.f.) and internal (i.e., those involving the electronic energy levels) \cite{PRAMasPelissonWolf}. As for the case of the van der Waals blockade, this approximation assumes implicitly that  the Casimir energy can be integrated out \emph{a priori} in the energy levels, before the atom is driven. However, dynamical effects may deviate from this assumption. Let us take as an example an atom driven through a lambda configuration [see Fig.\ref{Rabir}($a$)] in the presence of a material surface. On the one hand, the atom is pumped by two Raman lasers from two low-lying states, $|g\rangle$ and $|e\rangle$, to common excited intermediate states, $|i\rangle$. It is the combined action of both lasers that gives rise to the effective coherent oscillation of the atom between the states $|g\rangle$ and $|e\rangle$ \cite{QO,Eberly1987,JaynesCummings1963}.  On the other hand, the CP interaction of the atom with the surface  results in  similar excitation/desexcitation transitions which are mediated by virtual photons rather than laser beams --see Fig.\ref{Fig0}($d$).  Therefore, it is possible that both processes interfere with each other affecting the overall dynamics of the atom.  It is our purpose to investigate the conditions under which the interplay between the Raman interaction and the CP interaction provides appreciable consequences on the coherent dynamics of an atom. We will show that, in the perturbative regime, the net effect is an effective shift of the Rabi frequency.  

In this article we concentrate on scenarios similar to those of Refs.\cite{PRALambrechtetal,PRAMasPelissonWolf,chip}, where atoms are driven through a combination of Raman lasers and microwaves at the time they interact with a macroscopic surface. Our approach is based on  the time evolution of the atomic wave function.  As long as the surface resonant frequency is far from the atomic resonances, the classical treatment of the EM response of the surface is expected to be a good approximation.  The case of the interaction between two driven Rydberg atoms is left for a separate publication. 
The article is organized as follows. In Sec. \ref{first} we review the essentials of both the Rabi model and the Casimir interaction, and motivate our work. In Sec. \ref{Shifts} we address the renormalization of the energy levels and of the laser vertices, and show the role of the non-additive Casimir  terms on the coherent dynamics of a driven atom. We give explicit expressions for the effective shift of the Rabi frequency. Sec. \ref{Lambda} contains an explicit calculation of this shift for a $^{87}$Rb atom oscillating between two degenerate Zeeman sublevels close to a perfectly reflecting surface. In Sec. \ref{Ladder} the calculation is done for a $^{87}$Rb atom oscillating between two Rydberg states. In Sec. \ref{conclusions} we present our conclusions.
\begin{figure}[htb]
\includegraphics[height=6cm,width=8cm,clip]{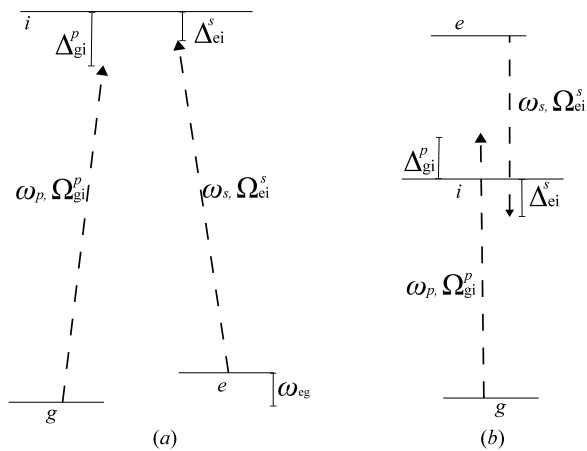}
\caption{$(a)$ Schematic representation of a lambda system, where two Raman lasers of frequencies $\omega_{p}$ and $\omega_{s}$  couple the states $|g\rangle$ and $|e\rangle$ to a third intermediate state $|i\rangle$ with coupling strengths $\Omega_{gi}^{p}$ and $\Omega_{ei}^{s}$ respectively. $(b)$ Schematic representation of a ladder system. In this case the intermediate state $|i\rangle$ lays between $|g\rangle$ and $|e\rangle$.}\label{Rabir}
\end{figure}

\section{Essentials of the EM interactions}\label{first}

We briefly review here the two EM interactions which govern the atomic dynamics. These are, the interaction with external electric fields and the interaction with the vacuum EM field.


\subsection{Laser fields interaction. Bare \footnote{Throughout this article we refer as bare to all those observables which are computed in the absence of surface induced quantum fluctuations.} Rabi oscillations}
Let us consider first the interaction of the atom with the external electric fields of two monochromatic lasers. An atom in free space with atomic levels $\{|i\rangle\}$ is described by the free Hamiltonian given by 
\begin{equation}
H_{0}^{at}=\sum_{i}\hbar\omega_{i}|i\rangle\langle i|,
\end{equation}
from which the unperturbed time-evolution operator reads, $\mathbb{U}^{at}(t)=\sum_{i}e^{-i\omega_{i}}|i\rangle\langle i|$. The Hamiltonian of interaction of the atom with the electric field of lasers of frequencies $\omega_{s}$ and $\omega_{p}$ is
\begin{align}
H^{int}_{ex}(t)&=\hbar\sum_{i}\Omega^{p}_{gi} \cos{\omega_{p}t}|i\rangle\langle g|+\Omega^{s}_{ei} \cos{\omega_{s}t}|i\rangle\langle e|\nonumber\\
+&\hbar\sum_{j\neq i,m}\Omega^{s}_{gj} \cos{\omega_{s}t}|j\rangle\langle g|+\hbar\sum_{m\neq i,j}\Omega^{p}_{em}\cos{\omega_{p}t}|m\rangle\langle g|\nonumber\\
+&\textrm{ h.c.}\label{HR}
\end{align}
where the strengths of the driven transitions are $\Omega^{p,s}_{gr,er}=-\langle g,e|\mathbf{d}|r\rangle\cdot\mathbf{E}_{p,s}$, with $\mathbf{d}$ the electric dipole moment operator and $\mathbf{E}_{p,s}$ the amplitude of the electric field of the lasers $p$ and $s$ at the position of the atom. The intermediate states labeled by $i$ are commonly accessible from $|g\rangle$ and $|e\rangle$. On the contrary, states labeled by $j$ are only accessible from $|g\rangle$ by the action of laser $s$ and  states labeled by $m$ are only accessible from $|e\rangle$ by the action of laser $p$. Adjusting conveniently the detuning of the lasers w.r.t. the transtion frequencies to the common states, with $\Delta^{p,s}_{gi,ei}=\omega_{ig,ie}-\omega_{p,s}$, as well as the strenghts of the transitions, it is possible to make the atom  oscillate coherently between
the states $|g\rangle$ and $|e\rangle$. That is for instance the case of an atom in either a lambda or a ladder system like those of Fig.\ref{Rabir}. Provided that  $|\Delta^{p,s}_{gi,ei}|\gg|\Omega^{p,s}_{gi,ei}|$ $\forall i$, the population of the common intermediate states can be eliminated adiabatically \footnote{The probability of excitation to the state $|i\rangle$ is proportional to $(\Omega^{p}_{gi}/\Delta^{p}_{gi})^{2}\sin^{2}{\Delta^{p}_{gi}t}+(\Omega^{s}_{ei}/\Delta^{s}_{ei})^{2}\sin^{2}{\Delta^{s}_{ei}t}$.} and the effective dynamics of the atom reduces to that of a two-level system \cite{Eberly1987}. Straightforward application of time-dependent perturbation theory with the interaction Hamiltonian of Eq.(\ref{HR}) yields the effective Rabi Hamiltonian  $H_{R}=H_{R}^{at}+H_{R}^{int}$, with
\begin{align}
 H_{R}^{at}&=\hbar(\omega_{g}+\delta\omega_{g})|g\rangle\langle g|+\hbar(\omega_{e}+\delta\omega_{e})|e\rangle\langle e|,\label{effi}\\
 H_{R}^{int}(t)&= \sum_{i}\frac{\hbar}{2}\Omega_{i}e^{i\omega_{L} t}|g\rangle\langle e|+h.c.,\label{effi2}
\end{align}
where $\delta\omega_{g,e}$  are the frequency light-shifts,  $\delta\omega_{g}=\sum_{i}(\Omega^{p}_{gi})^{2}/4\Delta^{p}_{gi}+\sum_{j\neq i}(\Omega^{s}_{gj})^{2}/4\Delta^{s}_{gj}$ and $\delta\omega_{e}=\sum_{i}(\Omega^{s}_{ei})^{2}/4\Delta^{s}_{ei}+\sum_{m\neq i}(\Omega^{p}_{em})^{2}/4\Delta^{p}_{em}$ respectively, $\Omega_{i}$ is the effective bare  Rabi frequency associated to the transition to the common state $|i\rangle$, $\Omega_{i}=\Omega^{p}_{gi}\Omega^{s}_{ei}(\Delta^{p}_{gi}+\Delta^{s}_{ei})/4\Delta^{p}_{gi}\Delta^{s}_{ei}$, and $\omega_{L}=\omega_{p}-\omega_{s}$ is the effective laser frequency. We note that for a 2-photon transition in ladder-configuration we must take $\omega_{s}<0$ in all the above equations. The diagrams contributing to the effective vertices of interaction and to the light-shifts are represented in Figs.\ref{Rabir}($a$) and ($b$) respectively. In particular, the equation for the effective vertex represented by the lower diagram of Fig.\ref{Rabir}($a$) reads
\begin{align}
H_{R}^{int}|_{ge}(t)&=\langle g|\mathbb{U}^{at}(t)\mathbb{U}^{\dagger}(t)H^{int}_{ex}(t)\mathbb{U}(t)\mathbb{U}^{at\dagger}(t)|e\rangle\nonumber\\
&\simeq\sum_{i}[\frac{-i}{2\hbar}\int_{0}^{t}\textrm{d}t' e^{-i\omega_{i}(t-t')}H^{int}_{ex}|_{ie}(t')e^{-i\omega_{e}t'}\nonumber\\
&\times H^{int}_{ex}|_{gi}(t)e^{i\omega_{g}t}]e^{i\omega_{eg}t}+[e\leftrightarrow g]^{\dagger}e^{i\omega_{eg}t},
\end{align}
where $\mathbb{U}(t)=$T-exp$\int_{0}^{t}\textrm{d}t'[H_{0}^{at}+H^{int}_{ex}(t')]$ and we have kept only the leading order terms in the second row. In Eq.(\ref{effi2}), far off-resonant terms and terms of order $\Delta^{s,p}_{ei,gi}/\omega_{ie,ig}$ smaller have been discarded.
\begin{figure}[htb]
\includegraphics[height=4.2cm,width=7.4cm,clip]{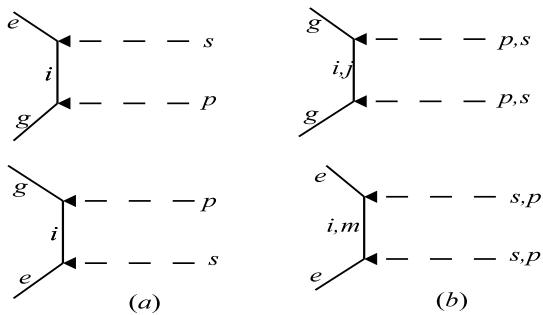}
\caption{$(a)$ Schematic representation of the effective vertices of the Rabi Hamiltonian, $H_{R}^{int}|_{eg}(t)$ and $H_{R}^{int}|_{ge}(t)$. $(b)$ Diagramatic representation of the light-shifts, $\delta\omega_{g}$ and $\delta\omega_{e}$.}\label{Rabir}
\end{figure}

Under the action of the Rabi Hamiltonian, an atom initially prepared in a linear combination of states $|g\rangle$ and $|e\rangle$ will undergo coherent Rabi oscillations. This problem is solved since long ago \cite{JaynesCummings1963}, for an extensive review see \cite{KnightRMP1980}. For the sake of comparison later-on we are interested in two particular cases for which analytical solutions are well-known. These are, that of equal laser frequencies in $\Lambda$-condiguration, $\omega_{L}=0$, and that for which the  rotating-wave-approximation (RWA) is applicable \footnote{The RWA  is valid as long as the effective laser frequency  is close to the atomic transition, $|\omega_{L}-\omega_{eg}|\ll\omega_{eg}$,  and much larger than the inverse time of observation, $\omega_{L}\gg T^{-1}$.}. The solution of the former is obtained by taking $\omega_{L}=0$ in the equations of the latter.

Transforming $H_{R}$ into the rotatory frame with rotation matrix $\mathbb{R}=-\omega_{L}|g\rangle\langle g|$ and discarding counter-rotating terms, the eigenstates and eigenenergies of the transformed Hamiltonian, $e^{i\mathbb{R}t}[H_{R}-\hbar\mathbb{R}]e^{-i\mathbb{R}t}$, read 
\begin{eqnarray}
|+\rangle&=&\cos{\theta_{c}}| g\rangle+\sin{\theta_{c}}|e\rangle,\label{+}\\
|-\rangle&=&\sin{\theta_{c}}| g\rangle-\cos{\theta_{c}}| e\rangle,\label{-}\\
E_{\pm}&=&(E_{ e}+E_{ g}+\hbar\omega_{L})/2\pm\frac{\hbar}{2}\Omega_{R},\label{E+-}
\end{eqnarray}
\begin{align}
\textrm{with }\cos{2\theta_{c}}&=\frac{\Delta}{\Omega_{R}},\textrm{ }
\sin{2\theta_{c}}=\frac{|\Omega|}{\Omega_{R}},\label{trigon}\\
\Delta&=\omega_{L}-\omega_{eg}-\delta\omega_{e}+\delta\omega_{g},\quad\label{globdelta}\\
\Omega&=\sum_{i}\Omega_{i},\quad\Omega_{R}=\sqrt{\Delta^{2}+|\Omega|^{2}}.\label{Rabfreq}
\end{align}
The wave function $\Psi(T)$ for a time $T>0$ of an atom driven by the above Hamiltonian and intially prepared at $t=0$ in a linear superposition of the states $|g\rangle$ and $|e\rangle$, $\Psi(0)=a_{g}(0)|g\rangle+a_{e}(0)|e\rangle$, is given in the Schr\"odinger picture by 
\begin{eqnarray}
\Psi(T)=\mathbb{U}^{R}(T)\Psi(0)&=&[ _{g}\textrm{U}^{R}_{g}(T)a_{g}(0)+\: _{g}\textrm{U}^{R}_{e}(T)a_{e}(0)]|g\rangle\nonumber\\
&+&[ _{e}\textrm{U}^{R}_{g}(T)a_{g}(0)+\: _{e}\textrm{U}^{R}_{e}(T)a_{e}(0)]|e\rangle,\nonumber
\end{eqnarray}
where the components $_{i}\textrm{U}^{R}_{j}(T)$, $i,j=g,e$, are given by
\begin{eqnarray}
_{g}\textrm{U}^{R}_{g}(T)&=&e^{-i(\omega_{g}-\Delta/2)T}[\cos{(\Omega_{R}T/2)}-i\frac{\Delta}{\Omega_{R}}\sin{(\Omega_{R}T/2)}],\nonumber\\
_{g}\textrm{U}^{R}_{e}(T)&=&-ie^{-i(\omega_{g}-\Delta/2)T}\frac{|\Omega|}{\Omega_{R}}\sin{(\Omega_{R}T/2)},\nonumber\\
_{e}\textrm{U}^{R}_{e}(T)&=&e^{-i(\omega_{e}+\Delta/2)T}[\cos{(\Omega_{R}T/2)}+i\frac{\Delta}{\Omega_{R}}\sin{(\Omega_{R}T/2)}],\nonumber\\
_{e}\textrm{U}^{R}_{g}(T)&=&-ie^{-i(\omega_{e}+\Delta/2)T}\frac{|\Omega|}{\Omega_{R}}\sin{(\Omega_{R}T/2)}.\label{UR}
\end{eqnarray}

\subsection{Vacuum field interaction}\label{vacuum}
We evaluate now the effect of the vacuum fluctuations on the dynamics of a free atom for the case that the state of the atom is a coherent superposition of the states $|g\rangle$ and $|e\rangle$ and the vacuum flucutations contain the interaction of free photons with a closeby dielectric surface.
In this respect, we use a semiclassical approach based on linear response theory. It consists of considering the photonic states as dressed by the classical interaction of free photons with the dielectric surface. This implies that the linear response of the EM field--i.e., its Green function, includes the scattering with the surface\footnote{The resultant EM interaction is also referred to in the literature as 'body assisted' interaction \cite{Safari}.}.  The interaction of dressed photons with the atom is treated quantum mechanically. As we did above for the driven atom, we restrict the calculation of  the time-evolution operator to the subspace $\{|g\rangle,|e\rangle\}$.

We constrain ourselves to the dipole approximation, $W=-\mathbf{d}\cdot\mathbf{E}(\mathbf{R})-\mathbf{m}\cdot\mathbf{B}(\mathbf{R})$, where $\mathbf{d}$ and $\mathbf{m}$ are the atomic electric and magnetic dipole operators respectively and $\mathbf{E}(\mathbf{R})$ and  $\mathbf{B}(\mathbf{R})$ are the electric and magnetic quantum field operators at the location of the atom, $\mathbf{R}$.
They can be decomposed in terms of $\omega$ modes as
\begin{eqnarray}
\mathbf{E}(\mathbf{R})&=&\int_{0}^{\infty}\textrm{d}\omega[\hat{\mathbf{E}}(\mathbf{R};\omega)+\hat{\mathbf{E}}^{\dagger}(\mathbf{R};\omega)],\nonumber\\
\mathbf{B}(\mathbf{R})&=&\int_{0}^{\infty}\textrm{d}\omega[\hat{\mathbf{B}}(\mathbf{R};\omega)+\hat{\mathbf{B}}^{\dagger}(\mathbf{R};\omega)],\nonumber
\end{eqnarray}
where $\hat{\mathbf{E}}(\mathbf{R};\omega)[\hat{\mathbf{B}}(\mathbf{R};\omega)]$ and $\hat{\mathbf{E}}^{\dagger}(\mathbf{R};\omega)[\hat{\mathbf{B}}^{\dagger}(\mathbf{R};\omega)]$ are the creation and annihilation electric (magnetic) field operators of photons
of energy $\hbar\omega$ at the position of the atom, $\mathbf{R}$.
The quadratic vacuum fluctuations of $\hat{\mathbf{E}}(\mathbf{R};\omega)$  satisfy the fluctuation-dissipation theorem (FDT) relations at zero temperature \cite{Agarwal1},
\begin{align}
\langle\tilde{0}|\hat{\mathbf{E}}(\mathbf{R};\omega)\otimes\hat{\mathbf{E}}^{\dagger}(\mathbf{R}';\omega)|\tilde{0}\rangle&=\frac{-\hbar\omega^{2}}{\pi\epsilon_{0}c^{2}}
\Im{[\mathbb{G}(\mathbf{R},\mathbf{R}';\omega)]},\label{FDT}\\
\langle\tilde{0}|\hat{\mathbf{B}}(\mathbf{R};\omega)\otimes\hat{\mathbf{B}}^{\dagger}(\mathbf{R}';\omega)|\tilde{0}\rangle&=\frac{\hbar}{\pi\epsilon_{0}c^{2}}
\Im{[\nabla_{R}\wedge\mathbb{G}(\mathbf{R},\mathbf{R}';\omega)\wedge\nabla_{R'}}],\nonumber
\end{align}
where  $|\tilde{0}\rangle$ is the EM vacuum state in the presence of the surface, $\Im$ denotes the imaginary part and $\mathbb{G}(\mathbf{R},\mathbf{R}';\omega)$ is the Green function of the Maxwell equation for the EM field,
\begin{eqnarray}
[\frac{\omega^{2}}{c^{2}}\mathbb{\epsilon}_{r}\cdot&-&\mathbf{\nabla}\wedge(\mathbb{\mu}_{r}^{-1}\cdot\mathbf{\nabla})\wedge]\mathbb{G}(\mathbf{R},\mathbf{R}';\omega)
=\delta^{(3)}(\mathbf{R},\mathbf{R}')\mathbb{I},\nonumber\\
Z,Z'&>&0.\label{Maxwelleq}
\end{eqnarray}
In this equation $\mathbf{\epsilon}_{r}$ and $\mathbf{\mu}_{r}$ are the relative electric permittivity and magnetic permeability tensors respectively, and the atom's position lies to the right of the surface, $Z>0$ [Fig.\ref{Fig0}($a$)].
The Green function can be decomposed into a free-space component and a scattering component. The contribution of the free space term to the Casimir energy is the ordinary free-space Lamb-shift that we consider included in the bare values of the atomic transition frequencies.

Next, we consider $W$ as a perturbation acting upon atomic and dressed photon states. The time-evolution operator of the latter is $\mathbb{U}_{0}(t)=[\mathbb{U}^{at}\otimes\mathbb{U}^{\gamma}](t)$, with $\mathbb{U}^{\gamma}(t)=\sum_{a=1}^{\infty}\prod_{j=1}^{a}\sum_{\gamma_{\mathbf{k}_{j},\mathbf{\epsilon}_{j}}}e^{-i\omega_{j} t}|\gamma_{\mathbf{k}_{j},\mathbf{\epsilon}_{j}}\rangle\langle \gamma_{\mathbf{k}_{j},\mathbf{\epsilon}_{j}}|$ the time-evolution operator of dressed multi-photon states of frequencies $\omega_{j}$, effective momentum $\hbar\mathbf{k}_{j}$ and polarization $\epsilon_{j}$. In the following , we will refer to $W$ as Casimir-Polder (CP) interaction.

We will show that for the case that the doublet $\{|g\rangle,|e\rangle\}$ is non-degenerate in comparison to the observation time,  $\omega_{eg}\gg T^{-1}$, the net effect of the vacuum fluctuations is an atomic level shift . On the contrary, for $\omega_{eg}\ll T^{-1}$ we will show that, in addition,  the vacuum fluctuations may induce Rabi oscillations in the degenerate doublet. In the latter case, the perturbative nature of the calculation is preserved as long as the CP interaction induced by the transition dipole moments $\langle g|\mathbf{d}|e\rangle$ and $\langle g|\mathbf{m}|e\rangle$ is negligible. We assume this condition in the following and apply  
 time-dependent perturbation theory \cite{Sakurai} for the calculation of the time-evolution operator projected on the subspace $|\Phi\rangle\equiv\{|e\rangle,|g\rangle\}\otimes|\tilde{0}\rangle$, 
\begin{eqnarray}
\mathbb{U}^{W}_{\Phi}(T)&=&[|\tilde{0}\rangle\langle\tilde{0}|\otimes(|g\rangle\langle g|+|e\rangle\langle e|)]\:\mathbb{U}_{0}(T)\label{Texp}\\
&\times&
\textrm{T-exp}\int_{0}^{T}\textrm{d}t\:\mathbb{U}_{0}^{\dagger}(t)\:W\:\mathbb{U}_{0}(t)\nonumber\\
&\times&[(|g\rangle\langle g|+|e\rangle\langle e|)\otimes|\tilde{0}\rangle\langle\tilde{0}|].\nonumber
\end{eqnarray}

\subsubsection{CP interaction in the non-degerate case. Atomic level shifts}

\begin{figure}[htb]
\includegraphics[height=4.5cm,width=8.4cm,clip]{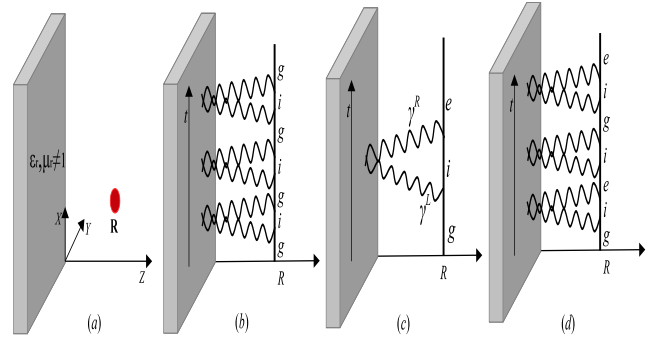}
\caption{ ($a$) Atom (in red) placed at $\mathbf{R}$ in front of a plane surface perpendicular to the $\hat{\mathbf{Z}}$ axis, with relative permittivity $\epsilon_{r}$ and permeability $\mu_{r}$.  ($b$) Feynman diagram of a multiple reflection process which contributes to $_{g}\textrm{U}^{W}_{g}$ in the non-degenerate case,
$\omega_{eg}T\gg1$. 
The diagram equals  three one-particle-irreducible additive phase shift factors, $(\delta E^{g}_{gg})^{3}$ --dissipative factors are omitted here for brevity. ($c$) Feynman diagram of a single reflection process which contributes to the non-additive shift factor, $\delta E^{g}_{ge}$.  Left-handed (L) and right-handed (R) photons are created/annihilated at the time the atom transits from $|g\rangle$ to $|i\rangle$ and from $|i\rangle$ to $|e\rangle$ respectively. ($d$) Feynman diagram of a multiple reflection process which contributes to $_{e}\textrm{U}^{W}_{g}$ in the degenerate case, $\omega_{eg}T\ll1$. The diagram equals  three one-particle-irreducible non-additive phase shift factors, $(\delta E^{\mathcal{E}}_{ge})^{2}\delta E^{\mathcal{E}}_{eg}$ --dissipative factors are omitted here for brevity.}\label{Fig0}
\end{figure}

In the non-degenerate case, with $\omega_{eg}T\gg1$, at any order in $T$ the diagrams which weight the most in the T-exponential of the diagonal components of $\mathbb{U}_{\Phi}^{W}$, $_{g}\textrm{U}^{W}_{g}$ and  $_{e}\textrm{U}^{W}_{e}$, are those in which the atom transits through intermediate virtual states before getting back repeatedly to the original state, $g$ and $e$ respectively. Each transition through intermediate states is accompanied by the emission and reabsorption of a single virtual photon which is reflected off the dielectric surface [see Fig.\ref{Fig0}($b$)]. 

On the contrary, the diagrams corresponding to the off-diagonal components, $_{g}\textrm{U}^{W}_{e}$ and  $_{e}\textrm{U}^{W}_{g}$, weight much less and the T-exponential can be truncated at leading order. That involves a single diagram in which the virtual photons are created at the atom in one of the states and annihilated at
the atom in the other state [see Fig.\ref{Fig0}($c$)]. Adding up the corresponding diagrams we obtain the  components,
\begin{eqnarray}
_{g}\textrm{U}^{W}_{g}(T)&=&e^{-i(\omega_{g}+\hbar^{-1}\delta E_{gg}^{g})T-\Gamma^{g}_{gg}T/2},\nonumber\\
_{g}\textrm{U}^{W}_{e}(T)&=&-e^{-i\omega_{g}T}[\delta E_{ge}^{e}-i\hbar\Gamma_{ge}^{g}/2]/E_{eg}\nonumber\\
&+&e^{-i\omega_{e}T}\delta E_{ge}^{e}/E_{eg},\nonumber\\
_{e}\textrm{U}^{W}_{e}(T)&=&e^{-i(\omega_{e}+\hbar^{-1}\delta E_{ee}^{e})T-\Gamma^{e}_{ee}T/2},\nonumber\\
_{e}\textrm{U}^{W}_{g}(T)&=&e^{-i\omega_{e}T}[\delta E_{eg}^{g}-i\hbar\Gamma_{eg}^{e}/2]/E_{eg}\nonumber\\
&-&e^{-i\omega_{g}T}\delta E_{eg}^{g}/E_{eg}.\nonumber
\end{eqnarray}
From these expressions we deduce that the off-diagonal components of $\mathbb{U}_{\Phi}^{W}$ can be neglected up to terms of order
$\mathcal{O}(\delta E_{eg,ge}^{g,e}/E_{eg})$
 for $\omega_{eg}T\gg1$. Thus, the effect of the CP interaction reduces here in good approximation to a renormalization of the energy levels of the atom. In particular,
$E_{g,e}\rightarrow E_{g,e}+\delta E_{gg,ee}^{g,e}-i\hbar\Gamma_{gg,ee}^{g,e}/2$.

In the above equations the additive energetic and dissipative terms have the usual expressions,
\begin{eqnarray}
\delta E_{gg}^{g}&=&-\sum_{ i,\gamma_{\mathbf{k},\mathbf{\epsilon}}}
\frac{|\langle i,\gamma|W| g,\tilde{0}\rangle|^{2}}{\hbar\omega+\hbar\omega_{ig}},\label{Eelggneqe}\\
\Gamma^{g}_{gg}&=&\frac{2\pi}{\hbar^{2}}\sum_{i,\gamma_{\mathbf{k},\mathbf{\epsilon}}}
\Theta(\omega_{gi})|\langle\gamma,i|W|g,\tilde{0}\rangle|^{2}\delta(\omega_{gi}-\omega).\label{gammag}
\end{eqnarray}
 Analogous expressions hold for $\delta E_{ee}^{e}$ and $\Gamma_{ee}^{e}$ with the substitution
$e\leftrightarrow g$.\\
\indent The non-additive energetic terms are,
\begin{eqnarray}
\delta E^{g}_{eg}&=&-\sum_{ i,\gamma_{\mathbf{k},\mathbf{\epsilon}}}
\frac{\langle e,\tilde{0}|W| i,\gamma\rangle\langle i,\gamma|W| g,\tilde{0}\rangle}{\hbar\omega+\hbar\omega_{ig}},\\\label{Egeg}
\delta E^{e}_{ge}&=&-\sum_{ i,\gamma_{\mathbf{k},\mathbf{\epsilon}}}
\frac{\langle g,\tilde{0}|W| i,\gamma\rangle\langle i,\gamma|W| e,\tilde{0}\rangle}{\hbar\omega+\hbar\omega_{ie}}.\label{Egee}
\end{eqnarray}
As for the non-additive dissipative terms, they are
\begin{eqnarray}
\Gamma^{g}_{ge}&=&\frac{2\pi}{\hbar^{2}}\sum_{i,\gamma_{\mathbf{k},\mathbf{\epsilon}}}\Theta(\omega_{gi})\langle g,\tilde{0}|W|\gamma,i\rangle
\langle\gamma,i|W|e,\tilde{0}\rangle\delta(\omega_{gi}-\omega),\nonumber
\end{eqnarray}
and an analogous expression holds for $\Gamma^{e}_{eg}$ with the substitution $e\leftrightarrow g$.
Single superscripts, $g$ or $e$, in the expressions for $\Gamma$ and $\delta E$  denote the reference frequency for the transitions within the sums. Double subscripts, $gg$, $ee$, $eg$ or $ge$, denote the bra and ket states in the quantum amplitudes.
In Appendix \ref{Appa} we evaluate the electric and magnetic vacuum field fluctuations which enter all the above quantities via the FDT.

\subsubsection{CP interaction in the degenerate case. CP induced Rabi oscillations}\label{vacuumRabi}

For $\omega_{eg}T\ll1$ we review the calculation on Ref.\cite{EPL} adding up some details to it.  In this case the diagrams which weight the most, both in the diagonal and in the off-diagonal components of $\mathbb{U}_{\Phi}^{W}$, are similar to those of Fig.\ref{Fig0}($b$) in the non-degenerate case, but for the fact that after each emission-reabsorpsion of a single virtual photon the atom may arrive at either state, $|g\rangle$ or $|e\rangle$, with a similar probability. Fig.\ref{Fig0}($c$) illustrates this process. Initially, a virtual photon of left-handed circular polarization, $\gamma^{L}$, is created at the time the atom is in state $|g\rangle$. Later, that photon is reflected off the surface turning into righ-handed circularly polarized, $\gamma^{R}$. Finally, the photon is annihilated at the time the atom gets to state $|e\rangle$. Diagrams involving $n$-photon intermediate states with $n>1$ can be disregarded in good approximation since their contribution is of the order of $(T\omega)^{1-n}$ 
times smaller than the diagram with $n$ single-photon intermediate states, with $\omega$ the typical frequency of the vacuum photons. In the non-retarded regime, $\omega\gg1/T$, so that multiphoton intermediate states are clearly negligible. In the retarded regime, $\omega\sim c/R$,  so that their neglect is possible for $T\gg R/c$, which is a realistic condition too. 
A typical diagram which contributes to the off-diagonal components of $\mathbb{U}_{\Phi}^{W}$ is depicted in Fig.\ref{Fig0}($d$). Their summation yields the following recurrent formulas,
\begin{eqnarray}
_{g}\textrm{U}^{W}_{g}(T)&=&e^{-i\omega_{\mathcal{E}}T}[1+\sum_{n=1}\delta^{(n)}_{\:\:\:\:g}\textrm{U}^{W}_{g}(T)],\\
_{e}\textrm{U}^{W}_{g}(T)&=&e^{-i\omega_{\mathcal{E}}T}\sum_{n=1}\delta^{(n)}_{\:\:\:\:e}\textrm{U}^{W}_{g}(T),\quad\textrm{ with}\nonumber\\
\delta^{(n)}_{\:\:\:\:g}\textrm{U}^{W}_{g}(T)&=&\frac{-iT(n-1)!}{\hbar n!}[(\delta E^{\mathcal{E}}_{gg}-i\hbar\Gamma^{\mathcal{E}}_{gg}/2)\delta^{(n-1)}_{\:\:\:\:\:\:\:\:\:\:g}\textrm{U}^{W}_{g}(T)\nonumber\\
&+&(\delta E^{\mathcal{E}}_{ge}-i\hbar\Gamma^{\mathcal{E}}_{ge}/2)\delta^{(n-1)}_{\:\:\:\:\:\:\:\:\:\:e}\textrm{U}^{W}_{g}(T)],\nonumber\\
\delta^{(n)}_{\:\:\:\: e}\textrm{U}^{W}_{g}(T)&=&\frac{-iT(n-1)!}{\hbar n!}[(\delta E^{\mathcal{E}}_{ee}-i\hbar\Gamma^{\mathcal{E}}_{ee}/2)\delta^{(n-1)}_{\:\:\:\:\:\:\:\:\:\:e}\textrm{U}^{W}_{g}(T)\nonumber\\
&+&(\delta E^{\mathcal{E}}_{eg}-i\hbar\Gamma^{\mathcal{E}}_{eg}/2)\delta^{(n-1)}_{\:\:\:\:\:\:\:\:\:\:g}\textrm{U}^{W}_{g}(T)],\nonumber\\
\delta^{(1)}_{\:\:\:\:g}\textrm{U}^{W}_{g}(T)&=&-iT\hbar^{-1}(\delta E^{\mathcal{E}}_{gg}-i\hbar\Gamma^{\mathcal{E}}_{gg}/2),\nonumber\\
\delta^{(1)}_{\:\:\:\:e}\textrm{U}^{W}_{g}(T)&=&-iT\hbar^{-1}(\delta E^{\mathcal{E}}_{eg}-i\hbar\Gamma^{\mathcal{E}}_{eg}/2),
\end{eqnarray}
where we have used $\mathcal{E}=(E_{g}+E_{e})/2$, $\omega_{\mathcal{E}}=\mathcal{E}/\hbar$. Analogous expressions hold for
$_{e}\textrm{U}^{W}_{e}(T)$ and $_{g}\textrm{U}^{W}_{e}(T)$ exchanging the subscripts $e\leftrightarrow g$ in the above equations.

From the diagram of Fig.\ref{Fig0}($d$) we observe that each factor $(\delta E^{\mathcal{E}}_{eg}-i\hbar\Gamma^{\mathcal{E}}_{eg}/2)$ flips the state
of the atom from $|e\rangle$ to $|g\rangle$, while each transposed factor produces an opposite flip. This is analogous to the action of the factors $\hbar\Omega/2$ and
$\hbar\Omega^{*}/2$ of the Rabi Hamiltonian respectively, except for the fact that here the damping terms break the time reversal symmetry. As a result, $\mathbb{U}_{\Phi}^{W}$ possesses the same functional form as $\mathbb{U}^{R}$ in Eq.(\ref{UR}), with the following substitution of the bare parameters in Eq.(\ref{UR}) with the tilded ones defined below \footnote{Note here
that $\tilde{\Omega}^{*}$ so defined is not the complex conjugate of $\tilde{\Omega}$ because of the damping factors.},
\begin{eqnarray}
\tilde{\omega}_{g}&\equiv&\omega_{g}+\delta E^{\mathcal{E}}_{gg}/\hbar-i\Gamma^{\mathcal{E}}_{gg}/2,\nonumber\\
\tilde{\omega}_{e}&\equiv&\omega_{e}+\delta E^{\mathcal{E}}_{ee}/\hbar-i\Gamma^{\mathcal{E}}_{ee}/2,\qquad
\tilde{\Delta}\equiv\tilde{\omega}_{e}-\tilde{\omega}_{g},\nonumber\\
\tilde{\Omega}&\equiv&2\delta E^{\mathcal{E}}_{ge}/\hbar-i\Gamma^{\mathcal{E}}_{ge},
\quad\tilde{\Omega}^{*}\equiv 2\delta E^{\mathcal{E}}_{eg}\hbar-i\Gamma^{\mathcal{E}}_{eg},\nonumber\\
|\tilde{\Omega}|^{2}&\equiv&\tilde{\Omega}\tilde{\Omega}^{*},\quad\Omega_{R}\equiv\sqrt{|\tilde{\Omega}|^{2}+\tilde{\Delta}^2}.\nonumber
\end{eqnarray}
This means that the Casimir-Polder interaction may indeed induce Rabi oscillations between two quasi-degenerate states \cite{EPL}.

\section{Vacuum-induced shift on the Rabi frequency}\label{Shifts}

In this section we derive an expression for the total shift induced on the Rabi frequency of a driven atom by vacuum fluctuations, $\delta\Omega_{R}=\delta_{1}\Omega_{R}+\delta_{2}\Omega_{R}+\delta_{3}\Omega_{R}$. In this equation we distinguish three different kinds of shifts, namely, that induced by the additive CP terms, $\delta_{1}\Omega_{R}$; the one induced by the non-additive CP terms, $\delta_{2}\Omega_{R}$; and the one induced by the renormalization of the vertices of interaction in $H^{int}_{ex}$, $\delta_{3}\Omega_{R}$.

\subsection{Shift by additive CP terms}
It is obvious that the additive terms of the Casimir interaction cause a shift on the atomic levels which induces a variation on the Rabi frequency \cite{Beguin}. First, the effective couplings $\Omega_{i}$ appearing in Eq.(\ref{effi}) experience a variation $\delta_{a}\Omega_{i}$ as a consequence of the shifts of the detunings. Second, the global detuning of Eq.(\ref{globdelta}), $\Delta$,  changes by an amount $\delta\Delta$ due to the shifts of the levels $E_{e}$ and $E_{g}$ as well as to the variations induced on the light-shifts, $\delta\omega_{e}$ and $\delta\omega_{g}$ [see Fig.\ref{Fig2}$(a)$]. As a result, as a function of the bare quantities $\Omega$,  $\Omega_{R}$ and $\Delta$, and of the additive CP terms,  we have at leading order
\begin{align}
\delta_{1}\Omega_{R}&\simeq\frac{\Omega}{\Omega_{R}}\sum_{i}\delta_{a}\Omega_{i}+\frac{\Delta}{\Omega_{R}}\delta\Delta\nonumber\\
&=-\sum_{i}\Bigl[\frac{\Omega\Omega_{gi}^{p}\Omega_{ei}^{s}}{4\hbar\Omega_{R}}[\frac{\delta E^{i}_{ii}-\delta E^{g}_{gg}}{\Delta^{p2}_{gi}}+\frac{\delta E^{i}_{ii}-\delta E^{e}_{ee}}{\Delta^{s2}_{ei}}]\nonumber\\&+\frac{\Delta}{4\hbar\Omega_{R}}[\frac{\Omega_{gi}^{p2}(\delta E^{i}_{ii}-\delta E^{g}_{gg})}{\Delta^{p2}_{gi}}-\frac{\Omega_{ei}^{s2}(\delta E^{i}_{ii}-\delta E^{e}_{ee})}{\Delta^{s2}_{ei}}]\Bigr]\nonumber\\&+\frac{\Delta}{\hbar\Omega_{R}}\Bigl[\delta E^{g}_{gg}-\delta E^{e}_{ee}-\sum_{j\neq i,m}\frac{\Omega_{gj}^{s2}}{4\Delta^{s2}_{gj}}(\delta E^{j}_{jj}-\delta E^{g}_{gg})\nonumber\\
&+\sum_{m\neq i,j}\frac{\Omega_{em}^{p2}}{4\Delta^{p2}_{em}}(\delta E^{m}_{mm}-\delta E^{e}_{ee})\Bigl].\label{addshift}
\end{align}

\subsection{Shift by non-additive CP terms}\label{Shift2}
\begin{figure}[htb]
\includegraphics[height=7.0cm,width=8.4cm]{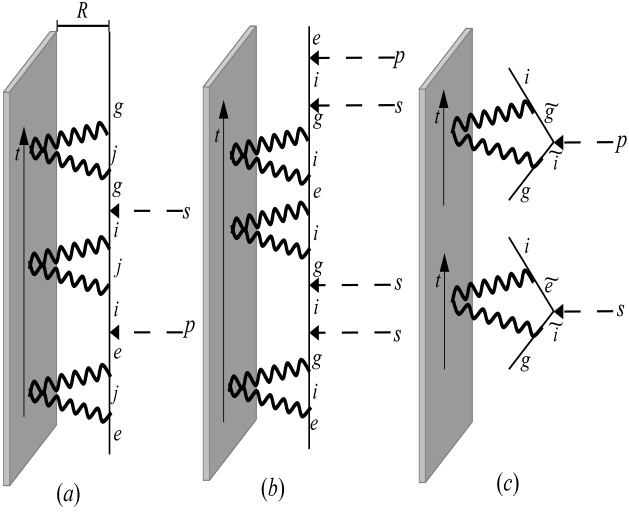}
\caption{ ($a$) Diagrammatic representation of a multiple reflection process in which additive CP terms induce a shift on the Rabi frequency of the kind $\delta_{1}\Omega_{R}$ of Eq.(\ref{addshift}).  One-loop-shifts in the levels $E_{g}$, $E_{e}$ and $E_{i}$ are included. $(b)$  Diagrammatic representation of a multiple reflection process in which non-additive CP terms induce a shift on the Rabi frequency of the kind $\delta_{2}\Omega_{R}$ of Eq.(\ref{nonaddshift}) for $\omega_{eg}\ll|\Omega|$. ($c$) Diagrammatic representation of the one-loop vertex renormalization processes which generates $\delta H^{p}_{ig}$ and $\delta H^{s}_{ig}$ respectively. They give rise to a shift on the Rabi frequency of the kind $\delta_{3}\Omega_{R}$ of Eq.(\ref{shift3}). The transient states $\tilde{i}$, $\tilde{g}$ and $\tilde{e}$ belong to the same energy levels as the states $i$, $g$ and $e$ respectively.}\label{Fig2}
\end{figure}

We have found in Sec. \ref{vacuumRabi} that for $\omega_{eg}T\ll1$ an induced Rabi frequency may be provided by vacuum fluctuations. In the case of a driven atom, $|\Omega|\sim1/T$, so that the degenerate condition is equivalent to the so-called \emph{deep strong coupling} (DSC) regime, $|\Omega|\gg\omega_{eg}$ \cite{Grimsmo}. The corresponding shift on $\Omega$ is given by diagrams in which pairs of consecutive Raman vertices and pairs of consecutive CP vertices alternate --eg. diagram of Fig.\ref{Fig2}($b$). As an example, we compute the leading order terms which contribute to the  variation of $_{g}\textrm{U}^{at}_{g}$, $\delta_{g}\textrm{U}_{g}$. To this aim,   
we treat $W+H^{int}_{R}$ as a perturbative interaction upon $H_{0}^{at}$.  Leading order contributions are of cubic order and correspond to the terms $H^{int}_{R}W^{2}$ and $W^{2}H^{int}_{R}$. At this order, we find
\begin{eqnarray}
\delta_{g}\textrm{U}_{g}&\simeq&-(iT/2E_{eg})\Bigl[e^{-i\omega_{g}T}\Omega(\delta E_{eg}^{e}-i\hbar\Gamma^{e}_{eg}/2)\label{2nddsc}\\&+&e^{-i\omega_{e}T}\Omega^{*}(\delta E_{ge}^{e}
-i\hbar\Gamma^{e}_{ge}/2)\Bigr]\quad\textrm{for }\omega_{eg}\gg|\Omega|,\nonumber\\
&\simeq&-\frac{T^{2}}{4\hbar}e^{-i\omega_{g}T}[\Omega(\delta E^{g}_{eg}-i\hbar\Gamma^{g}_{eg}/2)\label{1stdsc}\\
&+&\Omega^{*}(\delta E^{e}_{ge}-i\hbar\Gamma^{e}_{ge}/2)]\quad\textrm{for }\omega_{eg}\ll|\Omega|,\nonumber
\end{eqnarray}
where, in order to simplify matters, we consider zero global detuning, $\Delta\simeq0$, and we assume that all  the transition frequencies to intermediate states are much larger than $|\Omega|$. Straightforward comparison with the equation for $_{g}\textrm{U}^{R}_{g}$ reveals that Eq.(\ref{1stdsc}) is its term of $\mathcal{O}(T^{2})$ for 
\begin{eqnarray}
\Omega\rightarrow\Omega+2(\delta E^{g}_{ge}/\hbar-i\Gamma^{g}_{ge}/2),\nonumber\\
\Omega^{*}\rightarrow\Omega^{*}+2(\delta E^{e}_{eg}/\hbar-i\Gamma^{e}_{eg}/2).\label{renorm}
\end{eqnarray}

A similar calculation at order four in $W+H_{R}^{int}$ containing the terms $(H^{int}_{R})^{2}W^{2}$, $W^{2}(H^{int}_{R})^{2}$  and $H^{int}_{R}W^{2}H^{int}_{R}$, yields the following result,
\begin{eqnarray}
\delta_{g}\textrm{U}_{g}&=&e^{-i\omega_{g}T}\frac{i|\Omega|^{2}T^{3}}{12\hbar}[(\delta E_{gg}^{g}-i\hbar\Gamma_{gg}^{g}/2)\nonumber\\&+&(\delta E_{ee}^{e}-i\hbar\Gamma_{ee}^{e}/2)/2],\label{otri}
\end{eqnarray}
which holds for $|\Omega|$ either much larger or much smaller than $\omega_{eg}$. Again, comparison with the equation for $_{g}\textrm{U}^{R}_{g}$ reveals that Eq.(\ref{otri}) is its term of $\mathcal{O}(T^{3})$ for 
\begin{equation}\label{renorm2}
\Delta\rightarrow\Delta+[(\delta E_{gg}^{g}-\delta E_{ee}^{e})/\hbar-i(\Gamma_{gg}^{g}-\hbar\Gamma_{ee}^{e})/2].
\end{equation}
As expected, additive terms appear always as energy shifts irrespective of the magnitude of the ratio $|\Omega|/\omega_{eg}$. 

Disregarding dissipative terms, the shift of $\Delta$ was already accounted for in Eq.(\ref{addshift}). The additional CP induced shift on the Rabi frequency is therefore
\begin{equation}
\delta_{2}\Omega_{R}\simeq\frac{2\Omega}{\Omega_{R}}\Re{\{\delta E_{eg}\}}, \textrm{for }\omega_{eg}\ll|\Omega|.\label{nonaddshift}
\end{equation}

Alternatively, the shifts on the Rabi frequency induced by the additive and non-additive CP terms $\delta E^{\mathcal{E}}_{gg}$, $\delta E^{\mathcal{E}}_{ee}$, $\delta E^{\mathcal{E}}_{ge}$, $\delta E^{\mathcal{E}}_{eg}$ for $\omega_{eg}\ll|\Omega|$ can be computed out of the renormalization of the eigenstates $|+\rangle$, $|-\rangle$ and their corresponding eigenenergies --see Appendix\ref{app+-}.

\subsection{Shift by vertex renormalization}\label{Shift3}
Lastly, we are left with the effect of vacuum fluctuations on the Raman vertices appearing in Eq.(\ref{HR}). This corresponds to diagrams in which single Raman vertices and single CP vertices alternate. At leading order of time-dependent perturbation theory, the variations correspond to the diagrams of Fig.\ref{Fig2}($c$). They read
\begin{align}
\delta H^{int}_{ex}|_{ig}(t)&=\langle i,\tilde{0}|\mathbb{U}^{at}(t)\mathbb{U}^{W\dagger}(t)H^{int}_{ex}(t)\mathbb{U}^{W}(t)\mathbb{U}^{at\dagger}(t)|g,\tilde{0}\rangle\nonumber\\
&-\langle i|H_{ex}^{int}(t)|g\rangle\nonumber\\
&\simeq\hbar^{-2}\sum_{\gamma_{\omega},\tilde{i}}\int_{0}^{t}\textrm{d}t' e^{i(T-t')(\omega+\omega_{g})} e^{it'\omega_{i}}\langle i,\tilde{0}|W|\tilde{g},\gamma\rangle\nonumber\\
&\times e^{-i\omega_{ig}t}\langle \tilde{g},\gamma|H^{int}_{ex}(t)|\tilde{i},\gamma\rangle\nonumber\\
&\times\int_{0}^{t}\textrm{d}t'' e^{-i(T-t'')(\omega+\omega_{i})} e^{-it''\omega_{g}}\langle\tilde{i},\gamma|W|g,\tilde{0}\rangle\nonumber\\
&+\hbar^{-2}\sum_{\gamma_{\omega},\tilde{i}}\int_{0}^{t}\textrm{d}t' e^{i(T-t')(\omega+\omega_{e})} e^{it'\omega_{i}}\langle i,\tilde{0}|W|\tilde{e},\gamma\rangle\nonumber\\
&\times e^{-i\omega_{ig}t}\langle \tilde{e},\gamma|H^{int}_{ex}(t)|\tilde{i},\gamma\rangle\nonumber\\
&\times\int_{0}^{t}\textrm{d}t'' e^{-i(T-t'')(\omega+\omega_{i})} e^{-it''\omega_{g}}\langle \tilde{i},\gamma|W|g,\tilde{0}\rangle\nonumber\\
&\simeq\delta H^{p}_{ig}\cos{\omega_{p}t}+\delta H^{s}_{ig}\cos{\omega_{s}t},\label{hig}
\end{align}
with 
\begin{align}
\delta H^{p}_{ig}&=\sum_{\tilde{i},\tilde{g}}\frac{-\Omega_{\tilde{g}\tilde{i}}^{p}}{2\epsilon_{0}c^{2}}\Bigl[\omega_{ig}\textrm{Tr}\{\mathbf{d}_{i\tilde{g}}\cdot\mathbb{G}^{*}(\omega_{ig})\cdot\mathbf{d}_{\tilde{i}g}\}\nonumber\\
&+\frac{2}{\pi}\int_{0}^{\infty}\frac{\textrm{d}u\:u^{2}}{u^{2}+\omega_{ig}^{2}}\textrm{Tr}\{\mathbf{d}_{i\tilde{g}}\cdot\mathbb{G}(iu)\cdot\mathbf{d}_{\tilde{i}g}\}\Bigr],\nonumber\\
\delta H^{s}_{ig}&=\sum_{\tilde{i},\tilde{e}}\frac{-\Omega_{\tilde{e}\tilde{i}}^{s}}{\pi\epsilon_{0}c^{2}}\int_{0}^{\infty}\frac{\textrm{d}u\:u^{2}(u^{2}-\omega_{ig}\omega_{ei})}{(u^{2}+\omega_{ig}^{2})(u^{2}+\omega_{ei}^{2})}\nonumber\\
&\times\textrm{Tr}\{\mathbf{d}_{i\tilde{e}}\cdot\mathbb{G}(iu)\cdot\mathbf{d}_{\tilde{i}g}\}\nonumber,
\end{align}
where $\mathbf{d}_{ab}$ denotes $\langle a|\mathbf{d}|b\rangle$. 
An analogous equation for $\delta H^{int}_{ex}|_{ie}(t)$ yields
\begin{equation}\label{hie}
\delta H^{int}_{ex}|_{ie}(t)\simeq\delta H^{p}_{ie}\cos{\omega_{p}t}+\delta H^{s}_{ie}\cos{\omega_{s}t},
\end{equation}
with
\begin{align}
\delta H^{s}_{ie}&=\sum_{\tilde{i},\tilde{e}}\frac{-\Omega_{\tilde{e}\tilde{i}}^{s}}{2\epsilon_{0}c^{2}}\Bigl[\omega_{ei}\textrm{Tr}\{\mathbf{d}_{i\tilde{e}}\cdot\mathbb{G}^{*}(\omega_{ei})\cdot\mathbf{d}_{\tilde{i}e}\}\nonumber\\
&+\frac{2}{\pi}\int_{0}^{\infty}\frac{\textrm{d}u\:u^{2}}{u^{2}+\omega_{ie}^{2}}\textrm{Tr}\{\mathbf{d}_{i\tilde{e}}\cdot\mathbb{G}(iu)\cdot\mathbf{d}_{\tilde{i}e}\}\Bigr],\nonumber\\
\delta H^{p}_{ie}&=\sum_{\tilde{i},\tilde{g}}\frac{-\Omega_{\tilde{g}\tilde{i}}^{p}}{\epsilon_{0}c^{2}}\Bigl[\frac{\omega_{ig}^{2}}{\omega_{ig}-\omega_{ei}}\textrm{Tr}\{\mathbf{d}_{i\tilde{g}}\cdot\mathbb{G}^{*}(\omega_{ig})\cdot\mathbf{d}_{\tilde{i}e}\}\nonumber\\
&-\frac{\omega_{ei}^{2}}{\omega_{ig}-\omega_{ei}}\textrm{Tr}\{\mathbf{d}_{i\tilde{g}}\cdot\mathbb{G}(\omega_{ei})\cdot\mathbf{d}_{\tilde{i}e}\}\nonumber\\
&+\pi^{-1}\int_{0}^{\infty}\frac{\textrm{d}u\:u^{2}(u^{2}-\omega_{ig}\omega_{ei})}{(u^{2}+\omega_{ig}^{2})(u^{2}+\omega_{ei}^{2})}\nonumber\\
&\times\textrm{Tr}\{\mathbf{d}_{i\tilde{g}}\cdot\mathbb{G}(iu)\cdot\mathbf{d}_{\tilde{i}e}\}\Bigr]\nonumber.
\end{align}
In the above equations the tilded states $|\tilde{i}\rangle$,  $|\tilde{g}\rangle$ and $|\tilde{e}\rangle$ belong to the same energy levels as the states $i$, $g$ and $e$ respectively. For simplicity, we have assumed equal energies for all the intermediate states, $\omega_{\tilde{i}}=\omega_{i}$ $\forall \tilde{i}$.  Far off-resonant terms w.r.t. to the transition $|g\rangle\leftrightarrow |e\rangle$ and rapidly evanescent terms have been discarded. The complete expressions for $\delta H^{int}_{ex}|_{ig}(t)$ and  $\delta H^{int}_{ex}|_{ie}(t)$ can be found in Appendix \ref{AppVertex}.
We note that all the terms above are of the order of $\Omega^{p,s}_{ig,eg}\delta E^{g,e}_{gg,ee}/\omega_{ig,ei}$. Therefore, their contribution is generally relevant for very strong CP interaction w.r.t. the transition frequencies to intermediate states. This might be for instance the case of an atom which is made oscillate between two close Rydberg states in ladder-configuration, near a metallic surface. Thus, $\omega_{ei}>0$ has been assumed in the  above equations. 
It is now straightforward to calculate the effect of vertex renormalization (\emph{v}) on the coherent evolution of the atomic wave function. By comparing Eqs.(\ref{hig},\ref{hie}) with Eq.(\ref{HR}) we read  that it consists of a shift on the effective bare Rabi frequencies. Generally we find,
\begin{equation}
\delta_{v}\Omega_{i}=(\delta H^{p}_{gi}/\hbar\Omega^{p}_{gi}+\delta H^{s}_{ei}/\hbar\Omega^{s}_{ei})\Omega_{i},\label{Otilde1}
\end{equation}
while for the special case $|2\omega_{p,s}-\omega_{L}|\ll|\Delta|$ we have,
\begin{equation}
\delta_{v}\Omega_{i}=[(\delta H^{p}_{gi}+\delta H^{s}_{gi})/\hbar\Omega^{p}_{gi}+(\delta H^{s}_{ei}+\delta H^{p}_{ei})/\hbar\Omega^{s}_{ei}]\Omega_{i}.\label{Otilde2}  
\end{equation}
It is now straightforward to compute the total variation of the Rabi frequency due to vertex renormalization,
\begin{equation}\label{shift3}
\delta_{3}\Omega_{R}\simeq\frac{\Omega}{\Omega_{R}}\sum_{i}\delta_{v}\Omega_{i}.
\end{equation}

\section{CP induced shifts on the Rabi frequency of a rubidium atom close to a reflecting surface}\label{Rubidium}

\subsection{Oscillations between degenerate Zeeman sublevels in $\Lambda$-configuration}\label{Lambda} 

In Ref.\cite{EPL} it was found that a CP induced Rabi frequency as large as $2\pi$Hz could be obtained between the Zeeman sublevels  $|g\rangle=|5^{2}S_{1/2},F=1,m_{F}=-1\rangle$ and $|e\rangle=|5^{2}S_{1/2},F=1,m_{F}=+1\rangle$ of a $^{87}$Rb atom at zero temperature when it is placed in the vicinity of a perfectly reflecting surface parallel to the quantization axis, and as long as $|g\rangle$ and $|e\rangle$ remain degenerate, $\omega_{eg}\ll2\pi$Hz. Symmetry considerations imply that the CP interaction respects this condition since the additive energy shifts of both states  are equivalent. However, the presence of stray magnetic fields may induce a Zeeman splitting, $\hbar\omega_{eg}=\mu_{B}B_{0}$, with
$B_{0}$ the effective strength of the magentic fields along the quantization axis, say $\hat{\mathbf{X}}$, which could break the quasi-degeneracy of the states and violate the condition $\omega_{eg}\ll\tilde{\Omega}_{R}=2\delta E^{\mathcal{E}}_{eg}/\hbar$. In order to avoid this potential problem we propose an alternative setup in which we aim to measure the CP induced shift on the bare Rabi frequency of a driven atom in the DSC regime. The computation is in all points equivalent to the one of Ref.\cite{EPL}. The advantage of working with a driven atom is that, for a sufficiently high value of the bare Rabi frequency, $|\Omega|$, the unknown quantity $\omega_{eg}$ contributes to the actual Rabi frequency with a shift $\simeq \omega_{eg}^{2}/2|\Omega|$, which can be made negligible for sufficiently large $|\Omega|$ w.r.t. the shift provided by the non-additive CP terms, $2\delta E^{\mathcal{E}}_{eg}/\hbar$.

\begin{figure}[htb]
\includegraphics[height=10cm,width=8.2cm,clip]{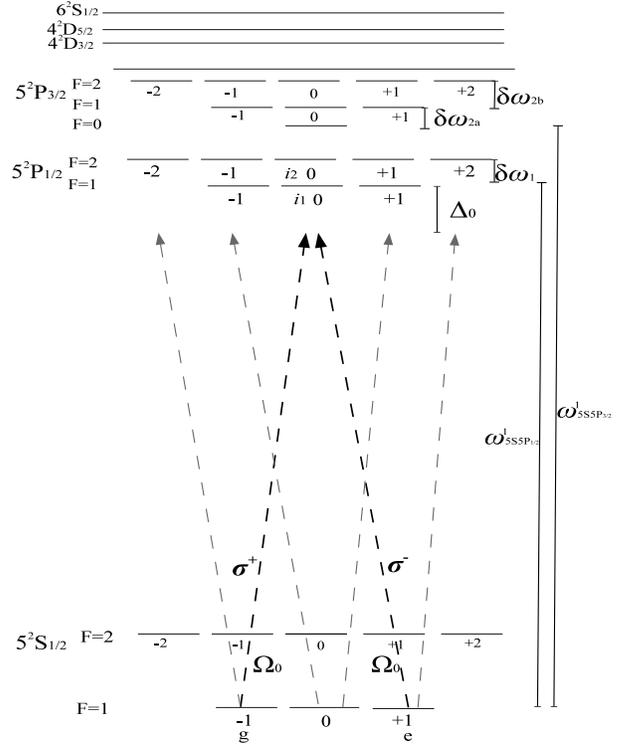}
\caption{Schematic representation of the action of two Raman lasers which drive a $^{87}$Rb atom between two Zeeman sublevels of the
ground state, $|g\rangle=|5^{2}S_{1/2},F=1,m_{F}=-1\rangle$ and $|e\rangle=|5^{2}S_{1/2},F=1,m_{F}=+1\rangle$. The quantization axis is
taken along the $\hat{\mathbf{X}}$ direction. The lasers have opposite polarization, $\sigma^{+}$ and $\sigma^{-}$, equal frequency, $\lambda\simeq795$nm such that $\Delta_{0}=2\pi10.0$GHz and equal intensity such that $\Omega_{0}=2\pi8.0$MHz.
They couple both states to a pair of common intermediate states of the level $|5^{2}P_{1/2}\rangle$, $| i_{1}\rangle=|5^{2}P_{1/2},F=1,m_{F}=0\rangle$ and $|i_{2}\rangle=|5^{2}P_{1/2},F=2,m_{F}=0\rangle$, with transition frequencies $\omega_{5S5P_{1/2}}^{1}$ and $\omega_{5S5P_{1/2}}^{1}+\delta\omega_{1}$ respectively, where $\omega_{5S5P_{1/2}}^{1}=2\pi377.107$THz and the hyperfine interval is $\delta\omega_{1}=2\pi0.817$GHz. Grey lines represent secondary couplings of the three Zeeman
sublelevels of the ground state, $|5^{2}S_{1/2},F=1\rangle$,  to other states in $|5^{2}P_{1/2}\rangle$. They are relevant for the computation of light-shifts. In addition to the D1 transition line, relevant to the calculation of $\delta_{2}\Omega_{R}$ are the D2 line transitions from  $|g\rangle$ and $|e\rangle$ to the  Zeeman sublevels of $5P_{3/2}$ with $m_{F}=0$. For those transtions, $\omega_{5S5P_{3/2}}^{1}=2\pi384.230$THz, $\delta\omega_{2a}=-2\pi0.072$GHz,  $\delta\omega_{2b}=2\pi0.157$GHz. Also, relevant to $\delta_{1}\Omega_{R}$ are the transitions $5P_{1/2}\leftrightarrow6S_{1/2}$ and $5P_{1/2}\leftrightarrow4D_{3/2}$.}\label{Raman}
\end{figure}

The setup is sketched in Fig.\ref{Raman},
where the parameters have been chosen so that the value of the effective bare Rabi frequency is  $\Omega_{R}=|\Omega|=2\pi20$Hz. Two Raman lasers of equal frequency and opposite circular polarization drive the transitions from $|g\rangle$ and $|e\rangle$ respectively to  a virtual state close to  $| i_{1}\rangle=|5^{2}P_{1/2},F=1,m_{F}=0\rangle$ and $|i_{2}\rangle=|5^{2}P_{1/2},F=2,m_{F}=0\rangle$. 
For convenience we take $\Delta^{s}_{gi_{1}}=\Delta^{p}_{ei_{1}}=\Delta_{0}=10$GHz, and the same intensity for both lasers, $|\Omega_{0}|=2\pi8.0$MHz, such that $|\Omega_{0}|\ll\Delta_{0}\ll\omega_{5S5P_{1/2}}^{1}$.
Denoting their electric field strength by $|\mathbf{E}_{0}|$, we define
\begin{equation}
\Omega_{0}\equiv\hbar^{-1}\langle 5^{2}S_{1/2}||\mathbf{d}||5^{2}P_{1/2}\rangle|\mathbf{E}_{0}|,
\end{equation}
and the effective bare Rabi frequency can be computed as a function of $\Omega_{0}$, $\Delta_{0}$ and $\delta\omega_{1}$ as,
\begin{eqnarray}
\Omega&=&|\mathbf{E}_{0}|^{2}\Bigl[\langle g|d_{+}|i_{1}\rangle\langle e|d_{-}|i_{1}\rangle^{*}/2\Delta_{0}\nonumber\\&+&\langle g|d_{+}|i_{2}\rangle
\langle e|d_{-}|i_{2}\rangle^{*}/2(\Delta_{0}+\delta\omega_{1})\Bigr]\nonumber\\
&\simeq&-\frac{|\Omega_{0}|^{2}}{24\Delta^{2}_{0}}\delta\omega_{1},\label{Omeq}
\end{eqnarray}
where the hyperfine interval satisfies $\delta\omega_{1}\ll\Delta_{0}$ and the dipole moment operator is expressed in the spherical basis.
The expectation values and transition frequencies have been taken from Ref.\cite{SeckRb87}. 

 \indent Beside the couplings between the states $|g\rangle$ and $|e\rangle$ and the common intermediate states $|i_{1}\rangle$ and $|i_{2}\rangle$, the Raman lasers also couple the Zeeman sublevels of the multiplet $|5^{2}S_{1/2},F=1\rangle$ to all the hyperfine Zeeman sublevels within $|5^{2}P_{1/2}\rangle$. The pairs of couplings are depicted in Fig.\ref{Raman} with gray lines. The result is that while the differential light-shift between $|e\rangle$ and $|g\rangle$ is exactly zero, there exists a differential light-shift between the state $|5^{2}S_{1/2},F=1,m_{F}=0\rangle$ and the other two states. This light-shift breaks the degeneracy between the three Zeeman subleves and  rises $|5^{2}S_{1/2},F=1,m_{F}=0\rangle$ above $|g\rangle$
 and $|e\rangle$ by an amount $\hbar|\Omega|/2=10$Hz$2\pi\hbar$. In turn, this is an advantage, since the states $|g\rangle$ and $|e\rangle$ become the lowest energy states and no dissipative decay terms enter in the calculation. Only at very short distances the positive magnetic energy shifts, $\mu_{B}^{2}/192\pi\epsilon_{0}c^{2}Z^{3}$, may take the states $|g\rangle$ and $|e\rangle$ over $|5^{2}S_{1/2},F=1,m_{F}=0\rangle$ \cite{EPL}. 
 
We proceed to compute $\delta\Omega_{R}$. In the first place, given that the bare global detuning is approximately zero, the Rabi shift induced by additive CP terms is caused by the variation of $\Delta_{0}$ in the equation for $\Omega$ [Eq.(\ref{Omeq})]. The variation of $\Delta_{0}$ is itself due to the differential level shift between the states $5S_{1/2}$ and $5P_{1/2}$. In the near field, $Z\lesssim100$nm, the main contribution to it comes from the difference between the reduced dipole matrix elements $\langle 6^{2}S_{1/2}||\mathbf{d}||5^{2}P_{1/2}\rangle$,  $\langle 4^{2}D_{3/2}||\mathbf{d}||5^{2}P_{1/2}\rangle$ and $\langle 5^{2}S_{1/2}||\mathbf{d}||5^{2}P_{1/2,3/2}\rangle$ \cite{Safronova}. As a result, we have 
$\delta_{1}\Omega_{R}\simeq\frac{2\Omega_{R}}{\Delta_{0}}(\delta E^{i}_{ii}-\delta E^{g}_{gg})$, 
where $i$ stands for any of the intermediate states, $i_{1}$ or $i_{2}$. Using the equations in Appendix \ref{Appa}  for the additive CP terms as functions of the Green's tensor which is, for a perfectly conducting reflector,
\begin{eqnarray}\label{Geeny}
G_{xx}(Z;\omega)&=&G_{yy}(Z;\omega)=\frac{e^{2ikZ}}{32\pi k^{2}Z^{3}}(-1+2ikZ+4k^{2}Z^{2}),\nonumber\\
G_{zz}(Z;\omega)&=&\frac{e^{2ikZ}}{16\pi k^{2}Z^{3}}(-1+2ikZ),
\end{eqnarray}
\begin{figure}[htb]
 \includegraphics[height=6cm,width=8.6cm,clip]{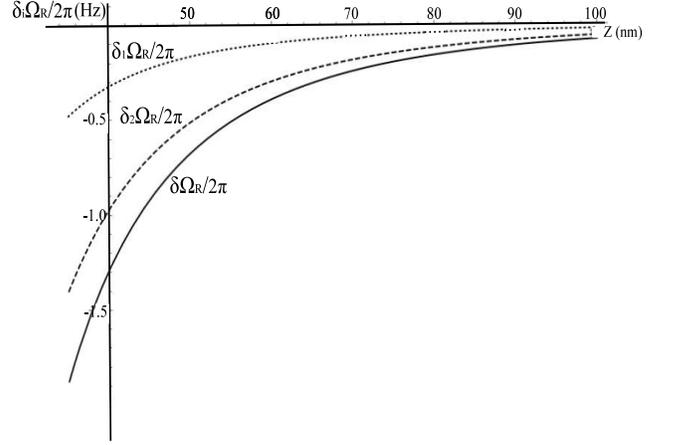}
\caption{Graphical representation of the Rabi frequency shifts as a function of the distance to the reflecting surface.}\label{Hy}
\end{figure}
the differential level shift in the near field can be written as 
\begin{align}
\delta E^{i}_{ii}-\delta E^{g}_{gg}&\simeq\frac{-1}{48\pi\epsilon_{0}Z^{3}}\Bigl[|\langle 6^{2}S_{1/2}||\mathbf{d}||5^{2}P_{1/2}\rangle|^{2}\\
&+\sum_{n=4}^{6}|\langle n^{2}D_{3/2}||\mathbf{d}||5^{2}P_{1/2}\rangle|^{2}\nonumber\\
&-|\langle 5^{2}S_{1/2}||\mathbf{d}||5^{2}P_{3/2}\rangle|^{2}
-2|\langle 5^{2}S_{1/2}||\mathbf{d}||5^{2}P_{1/2}\rangle|^{2}\Bigr].\nonumber
\end{align}



As for the Rabi shift induced by non-additive CP terms, the calculation was already carried out in Ref.\cite{EPL}. Here we just give the final formula,
\begin{eqnarray}
\delta_{2}\Omega_{R}&\simeq&\frac{1}{16\hbar c\pi^{2}\epsilon_{0}Z^{2}}\int_{0}^{\infty}\textrm{d}u\frac{f(u)}{u^{2}+\kappa_{jF}^{2}}\sum_{j=1/2,F=|j-3/2|}^{3/2,j+3/2}\omega^{F}_{5S5P_{j}}\nonumber\\
&\times&\langle e|d_{-}|5^{2}P_{j},F,m_{F}=0\rangle
\langle g|d_{+}|5^{2}P_{j},F,m_{F}=0\rangle,\nonumber
\end{eqnarray}
 where  $f(u)=e^{-2u}(1+2u-4u^{2})$ and $\kappa_{jF}=\omega^{F}_{5S5P_{j}}Z/c$, with  $\omega^{F}_{5S5P_{j}}$ being the transition frequency from the hyperfine level $F$ of  $5^{2}P_{j}$ to the states $|g\rangle$, $|e\rangle$.
 
 In Fig.\ref{Hy} we represent the values of the Rabi frequency shifts as a function of the distance to the surface in the non-retarded regime, $Z\lesssim100$nm.
It is clear that $\delta_{2}\Omega_{R}$ dominates over $\delta_{1}\Omega_{R}$ in our setup.

\subsection{Oscillations between Rydberg states in ladder-configuration}\label{Ladder}

\begin{figure}[htb]
 \includegraphics[height=7.2cm,width=4.9cm,clip]{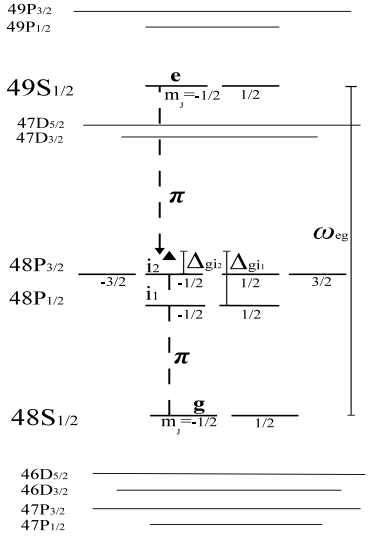}
\caption{Schematic representation of the action of a $\pi$-polarized microwave source which drives the oscillations of a $^{87}$Rb atom between 
two Rydberg levels,  $|g\rangle=|48^{2}S_{1/2},m_{j}=-1/2\rangle$ and $|e\rangle=|49^{2}S_{1/2},m_{j}=-1/2\rangle$. The quantization axis is
taken along the $\hat{\mathbf{X}}$ direction. The intermediate states are $|i_{1}\rangle=|48^{2}P_{1/2},m_{j}=-1/2\rangle$ and $|i_{2}\rangle=|48^{2}P_{3/2},m_{j}=-1/2\rangle$. Exact \emph{a priori} resonant condition is assumed, $\omega_{0}=\omega_{eg}/2=2\pi\:35.2387$GHz. Hence, $\Delta_{gi_{1,2}}=\Delta_{ei_{1,2}}$. 
}\label{Rydberg}
\end{figure}

Let us consider now the Rb atom close to the reflecting surface oscillating between two Rydberg states, $|g\rangle=|48^{2}S_{1/2},F=1,m_{j}=-1/2\rangle$ and
$|e\rangle=|49^{2}S_{1/2},F=1,m_{j}=-1/2\rangle$, under the action of a $\pi$-polarized microwave source which is resonant with the two-photon transition, 
$\omega_{0}=\omega_{eg}/2=2\pi\:35.2387$GHz [Fig.\ref{Rydberg}]. This setup is meant to mimic the situation of some hybrid quantum systems in which the coherent  
manipulation of Rydberg atoms is used to probe the EM fluctuations near a surface --eg. Ref.\cite{chip}. The microwave source couples the states $|g\rangle$ and $|e\rangle$ to the intermediate states 
$|i_{1}\rangle=|48^{2}P_{1/2},m_{j}=-1/2\rangle$ and
$|i_{2}\rangle=|48^{2}P_{3/2},m_{j}=-1/2\rangle$, with equivalent detunings $\Delta_{1}=\Delta_{i_{1}g,e}=-2\pi\:1.093$GHz, 
$\Delta_{2}=\Delta_{i_{2}g,e}=-2\pi\:0.1611$GHz, and with nearly equivalent strengths $-\Omega_{0}=\Omega_{i_{1,2}g}$, $\Omega_{i_{1,2}e}$. We set $\Omega_{0}=2\pi\:53.884$MHz
so that $\Omega_{R}=2\pi\:10.0$MHz. We note that, despite the two-photon resonance condition, there exists a bare global detuning due to the light-shift,
$\Delta=\delta\omega_{g}-\delta\omega_{e}=-2\pi\:1.719$MHz.

As already done in the previous section, the computaion of $\delta_{1}\Omega_{R}$ involves the evaluation of differential light-shifts. In the non-retarded regime this implies the addition and substraction of the squares of the reduced dipole matrix elements. While the $nS$ states couple only to $nP$ and $(n-1)P$ states in good approximation, $nP$ states couple also to $nD$, $(n-1)D$ and $(n-2)D$ states. This implies that the order of magnitude of the differential shifts in Eq.(\ref{addshift}) is $\sim\delta E_{48P48P}^{48P}$. Making use of the data in Refs.\cite{PRASP,Lietal} we obtain,
\begin{align}
\delta_{1}\Omega_{R}&\simeq\Bigl[\frac{\Omega}{\Omega_{R}}\frac{18.148}{Z^{3}(\mu\textrm{m})}+\frac{\Delta}{\Omega_{R}}\frac{8.147}{Z^{3}(\mu\textrm{m})}\Bigr]2\pi\:\textrm{MHz}\nonumber\\
&=-\frac{18.282}{Z^{3}(\mu\textrm{m})}2\pi\:\textrm{MHz}.
\end{align}

\begin{figure}[htb]
 \includegraphics[height=5.4cm,width=7.7cm,clip]{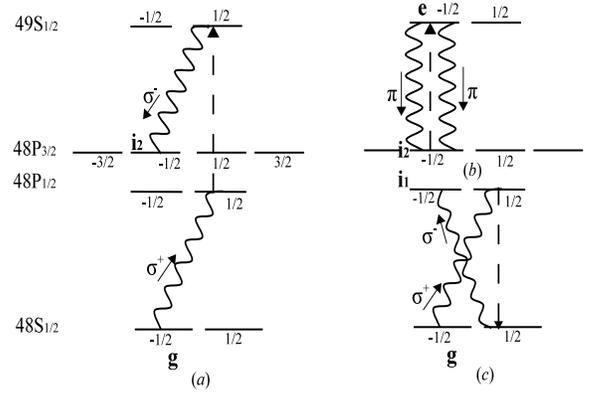}
\caption{Schematic representation of three processes which contribute to $\delta H^{s}_{gi_{2}}$, $(a)$;  $\delta H^{s}_{ei_{2}}$, $(b)$; and $\delta H^{p}_{gi_{1}}$, $(c)$, respectively. Wavy lines depict virtual transitions driven by vacuum photons. Dashed lines depict actual transitions driven by the microwave source. In the non-retarded regime $(a)$ is negligible and, after considering all kind of transitions, only processes of the kind of $(c)$ survive.
}\label{Rydberg2}
\end{figure}

As for the shift induced by the renormalization of the laser vertices, $\delta_{3}\Omega_{R}$, we use Eq.(\ref{shift3}) combined with Eq.(\ref{Otilde2})  since there exists only one microwave source. Nonetheless, it can be verified that, in the non-retarded regime, $\delta H^{s}_{i_{1,2}g}$ and $\delta H^{p}_{i_{1,2}e}$ are much smaller than $\delta H^{p}_{i_{1,2}g}$ and $\delta H^{s}_{i_{1,2}e}$, and hence negligible. This means that processes like that depicted in Fig.\ref{Rydberg2}$(a)$ in which two virtual transitions take place between $|g\rangle$ and $|e\rangle$ can be neglected, and we end up with $\delta_{v}\Omega_{i_{1,2}}\simeq-[\delta H^{p}_{i_{1,2}g} + \delta H^{s}_{i_{1,2}e}]\Omega_{i}/\Omega_{0}$. Discarding next the resonant components of $\delta H^{p}_{i_{1,2}g}$ and $\delta H^{s}_{i_{1,2}e}$, which are much smaller than the non-resonant components in the non-retarded regime, and using the Green's function of Eq.(\ref{Geeny})  in the near field, we obtain
\begin{align}
\delta H^{p}_{ig}&\simeq\frac{\Omega_{0}}{\omega_{0}64\pi\epsilon_{0}Z^{3}}\sum_{\tilde{i}\tilde{g}}[\langle i|d_{0}|\tilde{g}\rangle\langle\tilde{i}|d_{0}|g\rangle\nonumber\\
&-(3/2)\langle i|d_{-}|\tilde{g}\rangle\langle\tilde{i}|d_{+}|g\rangle].\label{Hpig}
\end{align} 
An analogous expression holds for $\delta H^{s}_{ie}$. In Eq.(\ref{Hpig}) conservation of total angular momentum implies that only processes mediated either by two $\pi$ virtual transtions or by two consecutive $\sigma_{+}$ and $\sigma_{-}$ transitions yield nonvanishing contribution. In Figs.\ref{Rydberg2}$(b)$ and $(c)$ we depict two of these processes. It turns out that the contribution of those processes involving $\pi$ virtual transitions vanishes for $E_{i_{1}}\simeq E_{i_{2}}$. Therefore, we are left only with processes of the kind of Fig.\ref{Rydberg2}$(c)$, 
\begin{align}
\delta H^{p}_{ig}&\simeq\frac{-3\Omega_{0}}{\omega_{0}128\pi\epsilon_{0}Z^{3}}\langle i|d_{-}|48S_{1/2},+1/2\rangle\\
&\times[\langle48P_{1/2},+1/2|d_{+}|g\rangle+\langle48P_{3/2},+1/2|d_{+}|g\rangle].\nonumber
\end{align}

\begin{figure}[htb]
 \includegraphics[height=6cm,width=8.6cm,clip]{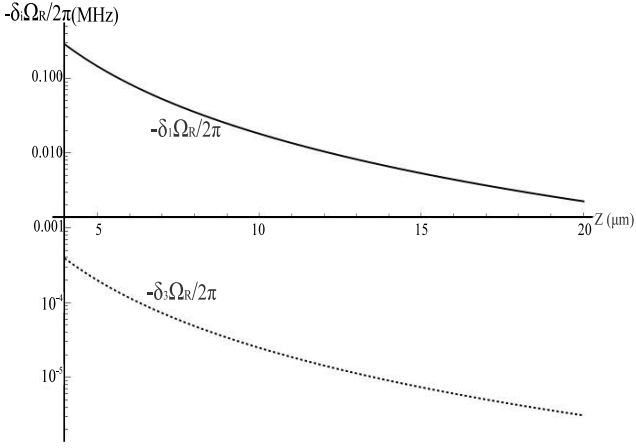}
\caption{Graphical representation of the Rabi frequency shifts as a function of the distance to the reflecting surface.}\label{Fig8}
\end{figure}

Finally, adding the contribution of $\delta H^{s}_{ie}$ and making use of the fact that the reduced dipole matrix elements between $S_{1/2}$ and $P_{j}$ states hardly depend on $j$ \cite{PRASP}, we can write in closed form,
\begin{align}
\delta_{3}\Omega_{R}&\simeq\frac{-\Omega}{\Omega_{R}}\frac{2\Omega_{i_{1}}+\Omega_{i_{2}}}{\omega_{0}512\pi\epsilon_{0}\hbar Z^{3}}[|\langle 48P||\mathbf{d}||48S\rangle|^{2}\nonumber\\
&+|\langle 48P||\mathbf{d}||49S\rangle|^{2}]\simeq-\frac{0.025}{Z^{3}(\mu\textrm{m})}2\pi\:\textrm{MHz},
\end{align}
which is almost three orders of magnitude smaller than $\delta_{1}\Omega_{R}$. A graphical representation is given in Fig.\ref{Fig8}. The ratio $\delta_{1}\Omega_{R}/\delta_{3}\Omega_{R}$ can be worked out from their expressions in Section \ref{Shifts}. We have,  $\delta_{1}\Omega_{R}/\delta_{3}\Omega_{R}\sim \Delta_{2}/A\omega_{0}$, with $A$ a numerical prefactor of order unity. We conclude that, altough $\delta_{3}\Omega_{R}$ is generally a few orders of magnitude smaller than the ordinary $\delta_{1}\Omega_{R}$, it may have an
effect on high precision measurements.

 We finalize this section with a comment on the setup of Ref.\cite{PRLPelisson}, where a $^{87}$Rb atom is made oscillate in ladder-configuration between two hyperfine-structure states, $|g\rangle=|5^{2}S_{1/2},F=1,m_{F}=0\rangle$ and $|e\rangle=|5^{2}S_{1/2},F=2,m_{F}=0\rangle$, with $\omega_{eg}\simeq 7$GHz. The two states are connected by an M1 transition. An analogous calculation to the one performed above yields $\delta_{3}\Omega_{R}\sim \Omega/100Z^{3}$(nm), which is negligible for operational distances larger than $100$nm.  

\section{Conclusions}\label{conclusions}

We have analysed all one-loop radiative corrections which contribute to the shift on the Rabi frequency of a driven atom close to a material surface. In addition to the shift induced by the ordinary additive variations on the atomic levels, $\delta_{1}\Omega_{R}$, two novel contributions have been reported. A shift induced by the non-additive Casimir-Polder terms, $\delta_{2}\Omega_{R}$, is found to  dominte when the atom is made oscillate between two degenerate Zeeman sublevels in lambda-configuration. A shift induced by the renormalization of the laser vertices of interaction, $\delta_{3}\Omega_{R}$, contributes at higher order than $\delta_{1}\Omega_{R}$ for an atom which is made oscillate between two Rydberg states in ladder-configuration.

\acknowledgments
We thank M.-P. Gorza, R. Guerout and A. Maury for fruitful discussions. Financial support from ANR-10-IDEX-0001-02-PSL and ANR-13-BS04--0003-02 is gratefully acknowledged.

\appendix
\section{Additive and non-additive CP terms}\label{Appa}
In this Appendix we compile the expressions for the energy shift  and dissipative CP terms, $\delta E$ and $\Gamma$ respectively, both additive and non-additive. As in Sec. \ref{vacuum}, the single superscripts in the expressions $\Gamma_{kl}^{j}$ and $\delta E_{kl}^{j}$  denotes the reference frequency for the transitions, $j=g,e,\mathcal{E}$;
while the double subscript denotes the bra and ket states in the quantum amplitudes, $kl=gg,ee,ge,eg$. In addition, we use the notation $\mathbf{d}_{ab}=\langle a|\mathbf{d}|b\rangle$, $\mathbf{m}_{ab}=\langle a|\mathbf{m}|b\rangle$. 
We apply the FDT outlined in Section \ref{vacuum} for the evaluation of vacuum field fluctuations at zero temperature. We separate electric and magnetic field contributions and, for the sake of simplicity, we assume that the surface possesses no chiral response,
\begin{eqnarray}
\delta E^{j}_{kl}&=&-\sum_{ i,\gamma_{\mathbf{k},\mathbf{\epsilon}}}[1-\delta_{\mathcal{E}j}(\delta_{ie}+\delta_{ig})]\Bigl[
\frac{\langle k,\tilde{0}|W_{el}| i,\gamma\rangle\langle i,\gamma|W_{el}| l,\tilde{0}\rangle}{\hbar\omega+\hbar\omega_{ij}}\nonumber\\
&+&\frac{\langle k,\tilde{0}|W_{m}| i,\gamma\rangle\langle i,\gamma|W_{m}| l,\tilde{0}\rangle}{\hbar\omega+\hbar\omega_{ij}}\Bigr]\nonumber\\
&=&\frac{1}{\pi\epsilon_{0}c^{2}}\mathbb{P}\int_{0}^{\infty}\textrm{d}\omega\:\omega^{2}\sum_{i}[1-\delta_{\mathcal{E}j}(\delta_{ie}+\delta_{ig})]\nonumber\\
&\times &\textrm{Tr}\frac{\mathbf{d}_{ki}\cdot
\Im{[\mathbb{G}(\mathbf{R},\mathbf{R};\omega)]}\cdot\mathbf{d}_{il}}{\omega+\omega_{ij}}\nonumber\\
&-&\frac{1}{\pi\epsilon_{0}c^{2}}\mathbb{P}\int_{0}^{\infty}\textrm{d}\omega\sum_{i}[1-\delta_{\mathcal{E}j}(\delta_{ie}+\delta_{ig})]\nonumber\\
&\times &\textrm{Tr}\frac{\mathbf{m}_{ki}\cdot
\Im{[\nabla_{R}\wedge\mathbb{G}(\mathbf{R},\mathbf{R};\omega)\wedge\nabla_{R}}]\cdot\mathbf{m}_{il}}{\omega+\omega_{ij}},\label{EgegAp}\\
\Gamma^{j}_{kl}&=&\frac{2\pi}{\hbar^{2}}\sum_{i,\gamma_{\mathbf{k},\mathbf{\epsilon}}}[1-\delta_{\mathcal{E}j}(\delta_{ie}+\delta_{ig})]\Theta(\omega_{ji})\nonumber\\&\times&\Bigl[\langle k,\tilde{0}|W_{el}|\gamma,i\rangle\langle\gamma,i|W_{el}|l,\tilde{0}\rangle\delta(\omega_{ji}-\omega)\nonumber\\
&+&\langle k,\tilde{0}|W_{m}|\gamma,i\rangle\langle\gamma,i|W_{m}|l,\tilde{0}\rangle\delta(\omega_{ji}-\omega\Bigr]\nonumber\\
&=&\frac{-2}{\hbar\epsilon_{0}c^{2}}\sum_{i}[1-\delta_{\mathcal{E}j}(1-\delta_{ie}-\delta_{ig})]
\Theta(\omega_{ji})\omega_{ji}^{2}\textrm{Tr}\{\mathbf{d}_{ki}\nonumber\\
&\cdot&\Im[\mathbb{G}(\mathbf{R},\mathbf{R};\omega_{ji})]\cdot\mathbf{d}_{il}\}\nonumber\\
&+&\frac{2}{\hbar\epsilon_{0}c^{2}}\sum_{i}[1-\delta_{\mathcal{E}j}(\delta_{ie}+\delta_{ig})]
\Theta(\omega_{ji})\textrm{Tr}\{\mathbf{m}_{ki}\nonumber\\
&\cdot&\Im[\nabla\wedge\mathbb{G}(\mathbf{R},\mathbf{R};\omega_{ji})\wedge\nabla]\cdot\mathbf{m}_{il}\}.\label{gammageAp}
\end{eqnarray}
The factor $[1-\delta_{\mathcal{E}j}(\delta_{ie}+\delta_{ig})]$ in these expressions accounts for the removal of the states $g$ and $e$ from the sums when the reference energy level is $\mathcal{E}=(E_{g}+E_{e})/2$ in the quasi-degenerate case. This ensures the perturbative nature of the calculation.\\
\indent In general,  in the energy shift terms  we can distinguish resonant ($r$) and off-resonant ($or$) components \cite{Wylie,Buhmann,PRADonaire2}. The resonant components account for the single poles of the integrand in Eq.(\ref{EgegAp}),
\begin{eqnarray}
\delta E^{j}_{kl}|_{r}&=&\frac{1}{\epsilon_{0}c^{2}}\sum_{i}[1-\delta_{\mathcal{E}j}(1-\delta_{ie}-\delta_{ig})]
\Theta(\omega_{ji})\omega_{ji}^{2}\nonumber\\&\times&\textrm{Tr}\{\mathbf{d}_{ki}
\cdot\Re[\mathbb{G}(\mathbf{R},\mathbf{R};\omega_{ji})]\cdot\mathbf{d}_{il}\}\nonumber\\
&-&\frac{1}{\epsilon_{0}c^{2}}\sum_{i}[1-\delta_{\mathcal{E}j}(\delta_{ie}+\delta_{ig})]
\Theta(\omega_{ji})\textrm{Tr}\{\mathbf{m}_{ki}\nonumber\\
&\cdot&\Re[\nabla\wedge\mathbb{G}(\mathbf{R},\mathbf{R};\omega_{ji})\wedge\nabla]\cdot\mathbf{m}_{il}\}.
\end{eqnarray}
For the off-resonant components, making use of the properties of the Green's functions, $\mathbb{G}(\mathbf{R},\mathbf{R};-\omega)=\mathbb{G}^{*}(\mathbf{R},\mathbf{R};\omega)$, $\omega^{2}\mathbb{G}(\mathbf{R},\mathbf{R};\omega)\rightarrow 0$ as $|\omega|\rightarrow\infty$, and employing standard integration techniques
in the complex plane \cite{Wylie} we find,
\begin{eqnarray}
\delta E^{j}_{kl}|_{or}&=&\frac{-\sum_{i}}{\pi\epsilon_{0}c^{2}}\int_{0}^{\infty}\textrm{d}u\:u^{2}\sum_{i}[1-\delta_{\mathcal{E}j}(\delta_{ie}+\delta_{ig})]\nonumber\\
&\times &\omega_{ij}\textrm{Tr}\frac{\mathbf{d}_{ki}\cdot
\mathbb{G}(\mathbf{R},\mathbf{R};iu)\cdot\mathbf{d}_{il}}{u^{2}+\omega^{2}_{ij}}\nonumber\\
&-&\frac{\sum_{i}}{\pi\epsilon_{0}c^{2}}\int_{0}^{\infty}\textrm{d}u\sum_{i}[1-\delta_{\mathcal{E}j}(\delta_{ie}+\delta_{ig})]\label{or}\\
&\times &\omega_{ij}\textrm{Tr}\frac{\mathbf{m}_{ki}\cdot
\Im{[\nabla_{R}\wedge\mathbb{G}(\mathbf{R},\mathbf{R};iu)\wedge\nabla_{R}}]\cdot\mathbf{m}_{il}}{u^{2}+\omega^{2}_{ij}}.\nonumber
\end{eqnarray}
Finally we note that these expressions can be formally rewritten as functions of the atomic polarizabilities using the appropriate definitions \cite{Safari}.

\section{Renormalization of eigenenergies and eigenstates for $\omega_{eg}\ll|\Omega|$.}\label{app+-}

We consider the eigenstates of the Hamiltonian  $H_{R}$ of Eqs.(\ref{effi},\ref{effi2}), $|+\rangle$, $|-\rangle$, as stationary states upon which $W$ acts as a stationary perturbation. This is  a good approximation for small effective laser frequency, $\omega_{L}T\ll1$, and for the case that the virtual transition between $|g\rangle$ and $|e\rangle$ are irrelevant in the CP interaction. Application of time-independent perturbation
theory at order $W^{2}$, up to $\mathcal{O}(\delta E\Delta^{2}/|\Omega|^{2})$, yields the energy shifts
\begin{eqnarray}
\delta E_{+}&=&-\sum_{ i\neq g,e,\gamma_{\mathbf{k},\mathbf{\epsilon}}}
\frac{|\langle i,\gamma|W| +\rangle|^{2}}{\hbar ck+E_{i}-E_{+}}-\sum_{\gamma_{\mathbf{k},\mathbf{\epsilon}}}
\frac{|\langle -,\gamma|W| +\rangle|^{2}}{\hbar ck-\hbar|\Omega|}\nonumber\\
&\simeq&\frac{1}{2}(\delta E_{gg}^{\mathcal{E}}+\delta E_{ee}^{\mathcal{E}})+\Re{\{\delta E_{eg}^{\mathcal{E}}\}}\nonumber\\
&+&\frac{\Delta}{2|\Omega|}(\delta E_{gg}^{\mathcal{E}}-\delta E_{ee}^{\mathcal{E}})-\frac{\Delta^{2}}{2|\Omega|^{2}}\Re{\{\delta E_{eg}^{\mathcal{E}}\}},\label{et1}\\
\delta E_{-}&=&-\sum_{ i\neq g,e,\gamma_{\mathbf{k},\mathbf{\epsilon}}}
\frac{|\langle i,\gamma|W| -\rangle|^{2}}{\hbar ck+E_{i}-E_{-}}-\sum_{\gamma_{\mathbf{k},\mathbf{\epsilon}}}
\frac{|\langle +,\gamma|W| -\rangle|^{2}}{\hbar ck+\hbar|\Omega|}\nonumber\\
&\simeq&\frac{1}{2}(\delta E_{gg}^{\mathcal{E}}+\delta E_{ee}^{\mathcal{E}})-\Re{\{\delta E_{eg}^{\mathcal{E}}\}}\nonumber\\
&-&\frac{\Delta}{2|\Omega|}(\delta E_{gg}^{\mathcal{E}}-\delta E_{ee}^{\mathcal{E}})+\frac{\Delta^{2}}{2|\Omega|^{2}}\Re{\{\delta E_{eg}^{\mathcal{E}}\}}.\label{et2}
\end{eqnarray}

As already explained in Section \ref{vacuum}, the double subscripts in the quantities $\delta E$,
$gg$, $ee$, $eg$ or $ge$, denote the bra and ket states in the quantum amplitudes, while the superscript $\mathcal{E}$ denotes the common reference frequency, $\omega_{\mathcal{E}}=(\omega_{g}+\omega_{e})/2$, for the intermediate atomic transitions involved in their calculations. We have assumed $|\Omega|\ll\omega_{ie},\omega_{ig},\omega_{i\mathcal{E}}$ $\forall i$ relevant in the sums over intermediate states, $i\neq g,e$, so that we can approximate
$\delta E_{eg,gg,ee}^{\mathcal{E}}\simeq\delta E_{eg,gg,ee}^{\mathcal{E}\pm|\Omega|/2}$. As anticipated in Sec. \ref{Shift2}, the net result of these energy shifts is a renormalization of the bare parameters which enter $\mathbb{U}^{R}$,
\begin{eqnarray}
\omega_{g}&\rightarrow&\tilde{\omega}_{g}=\omega_{g}+\delta E_{gg}^{\mathcal{E}}/\hbar,\qquad\omega_{e}\rightarrow\tilde{\omega}_{e}=\omega_{e}
+\delta E_{ee}^{\mathcal{E}}/\hbar,\nonumber\\
\Delta&\rightarrow&\tilde{\Delta}=\Delta-(\delta E^{\mathcal{E}}_{ee}-\delta E^{\mathcal{E}}_{gg})/\hbar,\label{Delta}\\
|\Omega|&\rightarrow&|\tilde{\Omega}|=|\Omega|+2\Re{\{\delta E_{eg}^{\mathcal{E}}\}},\label{Omega}\\
\Omega_{R}&\rightarrow&\tilde{\Omega}_{R}=\Omega_{R}+2\Re{\{\delta E_{eg}^{\mathcal{E}}\}}\label{OmegaR}\\
&+&\frac{\Delta}{|\Omega|}(\delta E_{gg}^{\mathcal{E}}-\delta E_{ee}^{\mathcal{E}})-\frac{\Delta^{2}}{|\Omega|^{2}}\Re{\{\delta E_{eg}^{\mathcal{E}}\}}.\nonumber
\end{eqnarray}

For the sake of completeness we compute the variations on $\mathbb{U}^{R}$ due to the interaction $W$, $\delta\mathbb{U}^{R}$. To this aim we calculate the wave function at time $T>0$ for the initial condition $\Psi(0)=|g\rangle$, 
\begin{eqnarray}
|\Psi(T)\rangle&=&[\mathbb{U}^{R}+\delta\mathbb{U}^{RW}](T)|g\rangle\\
&=&e^{-iT(E_{+}+\delta E_{+})}[\cos^{2}{\tilde{\theta}_{c}}|g\rangle+
\sin{2\tilde{\theta}_{c}}|e\rangle/2]\nonumber\\&+&
e^{-iT(E_{-}+\delta E_{-})}[\sin^{2}{\tilde{\theta}_{c}}|g\rangle-
\sin{2\tilde{\theta}_{c}}|e\rangle/2],\nonumber
\end{eqnarray}
where the renormalized (tilded) trigonometric functions are given by the expressions in Eq.(\ref{trigon}) but for the replacement of the bare parameters with the renormalized ones of Eqs.(\ref{Delta}-\ref{OmegaR}). In turn,
$[\mathbb{U}^{R}+\delta\mathbb{U}^{R}](T)$ presents the same functional form as
the operator $\mathbb{U}^{R}(T)$ in Eq.(\ref{UR}) with the replacement of the bare parameters by the renormalized ones. 
For the sake of simplicity  we choose $\Delta=-\omega_{eg}$, so that all exponential prefactors in front of the components of $\mathbb{U}^{R}(T)$ in Eq.(\ref{UR}) become $e^{-i\omega_{\mathcal{E}}T}$ with $\omega_{\mathcal{E}}=(\omega_{g}+\omega_{e})/2$.
We obtain, at leading order in the
energy shifts up to terms of the order of $\mathcal{O}(\Delta\delta E/\hbar|\Omega|^{2})$,
\begin{eqnarray}
\delta_{g}\textrm{U}^{R}_{g}(T)&\simeq & e^{-i\omega_{\mathcal{E}}T}\Bigl\{\cos{(|\Omega|T/2)}\label{qsdiagonal}\\
&-&i\frac{\delta E^{\mathcal{E}}_{gg}}{\hbar}[\cos{(|\Omega|T/2)}T/2
+|\Omega|^{-1}\sin{(|\Omega|T/2)}]\nonumber\\&-&\frac{T}{\hbar}\Re{\{\delta E^{\mathcal{E}}_{ge}\}}\sin{(|\Omega|T/2)}\nonumber\\
&-&i\frac{\delta E^{\mathcal{E}}_{ee}}{\hbar}[\cos{(|\Omega|T/2)}T/2-|\Omega|^{-1}\sin{(|\Omega|T/2)}]\Bigr\},\nonumber\\
\delta_{e}\textrm{U}^{R}_{g}(T)&\simeq &-e^{-i\omega_{\mathcal{E}}T}\Bigl\{i\sin{(|\Omega|T/2)}\label{qscross}\\
&+&(\delta E^{\mathcal{E}}_{ee}+\delta E^{\mathcal{E}}_{gg})\sin{(|\Omega|T/2)}T/2\hbar\nonumber\\&+&
\frac{i}{\hbar}\Re{\{\delta E^{\mathcal{E}}_{eg}\}}T\cos{(|\Omega|T/2)}\Bigr\}.\nonumber
\end{eqnarray}

\section{Vertex renormalization}\label{AppVertex}
We give the complete expressions for the one-loop vertex shifts, $\delta H^{int}_{ex}|_{ig}(t)$ and $\delta H^{int}_{ex}|_{ie}(t)$. We restrict ourselves to the electric dipole approximation and assume that the atom is driven in ladder-configuration, $\omega_{ig},\omega_{ei}>0$.
\begin{align}
\delta H^{int}_{ex}|_{ig}(t)&=\sum_{\tilde{i},\tilde{g}}\frac{-\Omega_{\tilde{g}\tilde{i}}^{p}}{4\epsilon_{0}c^{2}}
\Bigl[\omega_{ig}\textrm{Tr}\{\mathbf{d}_{i\tilde{g}}\cdot\mathbb{G}^{*}(\omega_{ig})\cdot\mathbf{d}_{\tilde{i}g}\}e^{i\omega_{p}t}\nonumber\\
+&\omega_{ig}\textrm{Tr}\{\mathbf{d}_{i\tilde{g}}\cdot\mathbb{G}(\omega_{ig})\cdot\mathbf{d}_{\tilde{i}g}\}e^{i(\omega_{p}-2\omega_{ig})t}\nonumber\\
&+\frac{2}{\pi}\int_{0}^{\infty}\frac{\textrm{d}u\:u^{2}}{u^{2}+\omega_{ig}^{2}}(e^{i\omega_{p}t}+e^{i(\omega_{p}-2\omega_{ig})t})\nonumber\\
&\times\textrm{Tr}\{\mathbf{d}_{i\tilde{g}}\cdot\mathbb{G}(iu)\cdot\mathbf{d}_{\tilde{i}g}\}\nonumber\\
&-\frac{2}{\pi}\int_{0}^{\infty}\frac{\textrm{d}u\:u^{2}e^{-uT}}{u^{2}+\omega_{ig}^{2}}e^{i(\omega_{p}-\omega_{ig})t}\nonumber\\
&\times\textrm{Tr}\{\mathbf{d}_{i\tilde{g}}\cdot[\mathbb{G}(iu)-\mathbb{G}(-iu)]\cdot\mathbf{d}_{\tilde{i}g}\}\Bigr]+[\omega_{p}\rightarrow-\omega_{p}]\nonumber\\
&+\sum_{\tilde{i},\tilde{e}}\Bigl[\frac{-\Omega_{\tilde{e}\tilde{i}}^{s}}{2\pi\epsilon_{0}c^{2}}\int_{0}^{\infty}
\frac{\textrm{d}u\:u^{2}(u^{2}-\omega_{ig}\omega_{ei})}{(u^{2}+\omega_{ig}^{2})(u^{2}+\omega_{ei}^{2})}\nonumber\\
&\times\textrm{Tr}\{\mathbf{d}_{i\tilde{e}}\cdot\mathbb{G}(iu)\cdot\mathbf{d}_{\tilde{i}g}\}(e^{i\omega_{s}t}+e^{i(\omega_{s}+\omega_{ei}-\omega_{ig})t})\nonumber\\
&+\frac{e^{i(\omega_{s}-\omega_{ig})t}}{2}\int_{0}^{\infty}\frac{\textrm{d}u\:u^{2}e^{-uT}}{(iu-\omega_{ig})(iu-\omega_{ei})}\nonumber\\
&\times\textrm{Tr}\{\mathbf{d}_{i\tilde{e}}\cdot[\mathbb{G}(iu)-\mathbb{G}(-iu)]\cdot\mathbf{d}_{\tilde{i}g}\}\nonumber\\
&+\frac{e^{i(\omega_{s}+\omega_{ei})t}}{2}\int_{0}^{\infty}\frac{\textrm{d}u\:u^{2}e^{-uT}}{(iu+\omega_{ig})(iu+\omega_{ei})}\nonumber\\
&\times\textrm{Tr}\{\mathbf{d}_{i\tilde{e}}\cdot[\mathbb{G}(iu)-\mathbb{G}(-iu)]\cdot\mathbf{d}_{\tilde{i}g}\}\Bigr]+[\omega_{s}\rightarrow-\omega_{s}],\nonumber
\end{align}
\begin{align}
\delta H^{int}_{ex}|_{ie}(t)&=\sum_{\tilde{i},\tilde{e}}\frac{-\Omega_{\tilde{e}\tilde{i}}^{s}}{4\epsilon_{0}c^{2}}\Bigl[\omega_{ie}\textrm{Tr}\{\mathbf{d}_{i\tilde{e}}\cdot\mathbb{G}^{*}(\omega_{ie})\cdot\mathbf{d}_{\tilde{i}e}\}e^{i\omega_{s}t}\nonumber\\
&+\omega_{ie}\textrm{Tr}\{\mathbf{d}_{i\tilde{e}}\cdot\mathbb{G}(\omega_{ie})\cdot\mathbf{d}_{\tilde{i}e}\}e^{i(\omega_{s}-2\omega_{ie})t}\nonumber\\&+\frac{2}{\pi}\int_{0}^{\infty}\frac{\textrm{d}u\:u^{2}}{u^{2}+\omega_{ie}^{2}}(e^{i\omega_{s}t}+e^{i(\omega_{s}-2\omega_{ie})t})\nonumber\\&\times\textrm{Tr}\{\mathbf{d}_{i\tilde{e}}\cdot\mathbb{G}(iu)\cdot\mathbf{d}_{\tilde{i}e}\}\nonumber\\
&-\frac{2}{\pi}\int_{0}^{\infty}\frac{\textrm{d}u\:u^{2}e^{-uT}}{u^{2}+\omega_{ie}^{2}}e^{i(\omega_{s}-\omega_{ie})t}\nonumber\\&\times\textrm{Tr}\{\mathbf{d}_{i\tilde{e}}\cdot[\mathbb{G}(iu)-\mathbb{G}(-iu)]\cdot\mathbf{d}_{\tilde{i}e}\}\Bigr]+[\omega_{s}\rightarrow-\omega_{s}]\nonumber\\
&+\sum_{\tilde{i},\tilde{g}}\Bigl[\frac{-\Omega_{\tilde{g}\tilde{i}}^{p}}{2\pi\epsilon_{0}c^{2}}\int_{0}^{\infty}\frac{\textrm{d}u\:u^{2}(u^{2}-\omega_{ig}\omega_{ei})}{(u^{2}+\omega_{ig}^{2})(u^{2}+\omega_{ei}^{2})}\nonumber\\&\times\textrm{Tr}\{\mathbf{d}_{i\tilde{g}}\cdot\mathbb{G}(iu)\cdot\mathbf{d}_{\tilde{i}e}\}(e^{i\omega_{p}t}+e^{i(\omega_{p}+\omega_{ei}-\omega_{ig})t})\nonumber\\
&+\int_{0}^{\infty}\frac{\textrm{d}u\:u^{2}e^{-uT}}{(iu+\omega_{ig})(iu+\omega_{ei})}\nonumber\\&\times\textrm{Tr}\{\mathbf{d}_{i\tilde{g}}\cdot[\mathbb{G}(iu)-\mathbb{G}(-iu)]\cdot\mathbf{d}_{\tilde{i}e}\}\nonumber\\
&\times[e^{i(\omega_{p}-\omega_{ig})t}+e^{i(\omega_{p}+\omega_{ei})t}]/2\nonumber\\
&+\frac{\pi}{\omega_{ig}-\omega_{ei}}\textrm{Tr}\{\mathbf{d}_{i\tilde{g}}\cdot[\omega_{ig}^{2}\mathbb{G}(\omega_{ig})e^{i(\omega_{p}+\omega_{ei}-\omega_{ig})t}\nonumber\\
&+\omega_{ig}^{2}\mathbb{G}^{*}(\omega_{ig})e^{i\omega_{p}t}-\omega_{ei}^{2}\mathbb{G}^{*}(\omega_{ei})e^{i(\omega_{p}+\omega_{ei}-\omega_{ig})t}\nonumber\\
&-\omega_{ei}^{2}\mathbb{G}(\omega_{ei})e^{i\omega_{p}t}]\cdot\mathbf{d}_{\tilde{i}e}\}\Bigl]+[\omega_{p}\rightarrow-\omega_{p}].\nonumber
\end{align}
As in Sec. \ref{Shift3}, the tilded states $|\tilde{i}\rangle$,  $|\tilde{g}\rangle$ and $|\tilde{e}\rangle$ belong to the same energy levels as the states $i$, $g$ and $e$ respectively. Equal energies have been assumed for all the intermediate states, $\omega_{\tilde{i}}=\omega_{i}$ $\forall \tilde{i}$.

\end{document}